\documentclass[11pt,british,english,australian]{article}
\usepackage[T1]{fontenc}
\usepackage[latin9]{inputenc}
\usepackage{geometry}
\usepackage{amstext}
\usepackage{graphicx}
\usepackage[authoryear]{natbib}
\usepackage{babel}
\begin{document}

\title{Large large-trader activity weakens the long memory of limit order
markets}

\author{Kevin Primicerio$^{1}$ and Damien Challet$^{1,2}$\\
~\\
\foreignlanguage{british}{{\footnotesize{}$^{1}$~Mathématiques et
Informatique pour la Complexité et les Systèmes }}\\
\foreignlanguage{british}{{\footnotesize{}CentraleSupélec, Université
Paris-Saclay, 91190, Gif-Sur-Yvette, France}}\\
\foreignlanguage{british}{{\footnotesize{}~}}\\
\foreignlanguage{english}{{\footnotesize{}$^{2}\,$Encelade Capital
SA, EPFL Innovation Park, Building C}}\\
\foreignlanguage{english}{{\footnotesize{}1015 Lausanne, Switzerland}}}
\maketitle
\begin{abstract}
Using more than 6.7 billions of trades, we explore how the tick-by-tick
dynamics of limit order books depends on the aggregate actions of
large investment funds on a much larger (quarterly) timescale. In
particular, we find that the well-established long memory of market
order signs is markedly weaker when large investment funds trade either
in a directional way and even weaker when their aggregate participation
ratio is large. Conversely, we investigate to what respect a weaker
memory of market order signs predicts that an asset is being actively
traded by large funds. Theoretical arguments suggest two simple mechanisms
that contribute to the observed effect: a larger number of active
meta-orders and a modification of the distribution of size of meta-orders.
Empirical evidence suggests that the number of active meta-orders
is the most important contributor to the loss of market order sign
memory.
\end{abstract}

\section{Introduction}

Financial market dynamics is complex in part because of the very large
variety of timescales at play. Both traders and volatility feedback
loops are known to have widely distributed timescales \citep{zumbachlynch,lillo2007limit,zhou2011investment,tumminello2011identification,challet2016trader}.
Accordingly, investigating how timescales interact reveals some of
the fundamental dynamical ingredients of price dynamics. For example,
the asymmetric relationship between historical and realized volatility
shows that price dynamics is not symmetric with respect to time reversal
\citep{zumbachlynch,zumbach2009time}, which imposes a strong constraint
on realistic stochastic volatility models \citep{blanc2017quadratic}.

The long memory of the signs of market orders is a well-established
stylized fact of limit order books \citep{bouchaud2004fluctuations,lillo2004long}
that passes the most stringent statistical tests. \citet{lillo2005theory}
propose a mathematical framework that links the long memory of these
signs to the way very large orders are split into a series of smaller
market orders (thereby creating a meta-order) and is able to reproduce
the empirical auto-correlation function if the distribution of the
meta-order size has a Pareto-like tail. In other words, the shape
of the sign auto-correlation function reflects that of the distribution
of the size of meta-orders.

Here, we use two large databases of almost maximally different timescales,
namely quarterly filings by large investment funds and a comprehensive
tick-by-tick database, which allow us to investigate the influence
of large funds on the memory properties of the limit order book. We
first show that the memory length of market order signs (buy/sell)
of a given asset is markedly weaker when a large fraction of its capitalization
is exchanged by large funds over a quarter. Reciprocally, we test
if assets with the weakest market order sign memory are likely to
being much traded by large funds. Finally, we use the theoretical
framework of \citet{lillo2005theory} to put forward a coherent picture
of our findings. 

\section{The data}

Our dataset consists of two databases: quarterly snapshots of large
investment ownership, from the corresponding FactSet database in the
2007-2013 period (32 reports), which contains data 10845 funds with
more than USD 100 millions under management. We filter out funds with
less than USD 100'000 invested into securities. The remaining funds
are invested in 12531 securities. We focus on the 2480 assets continuously
recorded in FactSet database after their first quarter of appearance
and with at least one full year of record. Using automated methods,
we link assets found in both FactSet and the Thomson Reuters Tick
History databases. The latter provides an event-by-event history of
limit order books. For each asset traded on the NASDAQ and each day,
we extracted all the trade prices together with the best bid and ask
prices just before the trades. Finally, we keep assets traded for
at least 200 days and with more than 200 trades per day on average.
This leaves 846 stocks and more than 6.7 billion trades.

\section{Methods}

In order to link trade-by-trade data with quarterly fund filings,
we define suitable quantities in each dataset and investigate how
they are related. For each asset $\alpha$, we compute the mid-price
just before the $n$-th trade, denoted by $m_{\alpha,n}$ as the average
between the best bid and ask prices. Then we define the sign of the
$n$-th market order of asset $\alpha$ as
\[
\epsilon_{\alpha,n}=\text{sign}(p_{\alpha,n}-m_{\alpha,n}).
\]

We drop trades that occur exactly at the previous mid-price. As suggested
by the above notation, we define the time as the number of market
orders since the beginning of the time-series. 

\subsection{Microstructure: memory length of market order sign auto-correlation}

We define several simple ways to characterize the memory of market
order signs. Unless specified otherwise, all market microstructure
quantities are measured over a full trading week. The first one consists
in measuring the probability of the occurrence of $\kappa$ consecutive
trades of a given sign in a random contiguous subset of $\{s_{n}\}$.
Mathematically, for a generic $\kappa$, this amounts to measuring
the conditional event frequency
\begin{equation}
\pi_{\alpha}^{(s\kappa)}=P(s_{n}=s_{n+1}=\cdots=s_{n+\kappa}=s)\label{eq:pi_kappa}
\end{equation}
 for both $s\in\{-1,1\}$. 

Another way to characterize the memory of order signs is the market
order sign autocorrelation at lag $\tau$ (in unit of market orders),
denoted by $C_{\alpha}(\tau)$. Process $\epsilon_{n}^{\alpha}$ has
a long memory if the integral of its autocorrelation function $C_{\alpha}$
diverges. Many references find that $C_{\alpha}(\tau)\propto a\tau^{-b}$
where $b<1$, in which case the integral of $C_{\alpha}(\tau)$ is
infinite (see e.g. \citet{lillo2005theory,toth2015equity}) (we omit
the $\alpha$ index for $a$ and\textbf{ $b$} in order to avoid too
heavy notations). This is indeed a good approximation for very long
time series. For finite time series of length $N$, one can define
the effective memory length as the lag $\tau_{\alpha}^{*}$ after
which $C_{\alpha}$ reaches for the first time the noise level of
autocorrelation functions $2/\sqrt{N}$, i.e., $\tau_{\alpha}^{*}$
is such that $C_{\alpha}(\tau)>2/\sqrt{N}\,\,\,\forall\tau\le\tau_{\alpha}^{*}$
. We will also consider the scaled maximum lag $\tau_{\alpha}^{*}/N$.

\begin{figure}
\begin{centering}
\begin{minipage}[t]{0.49\columnwidth}%
\begin{center}
\includegraphics[scale=0.4]{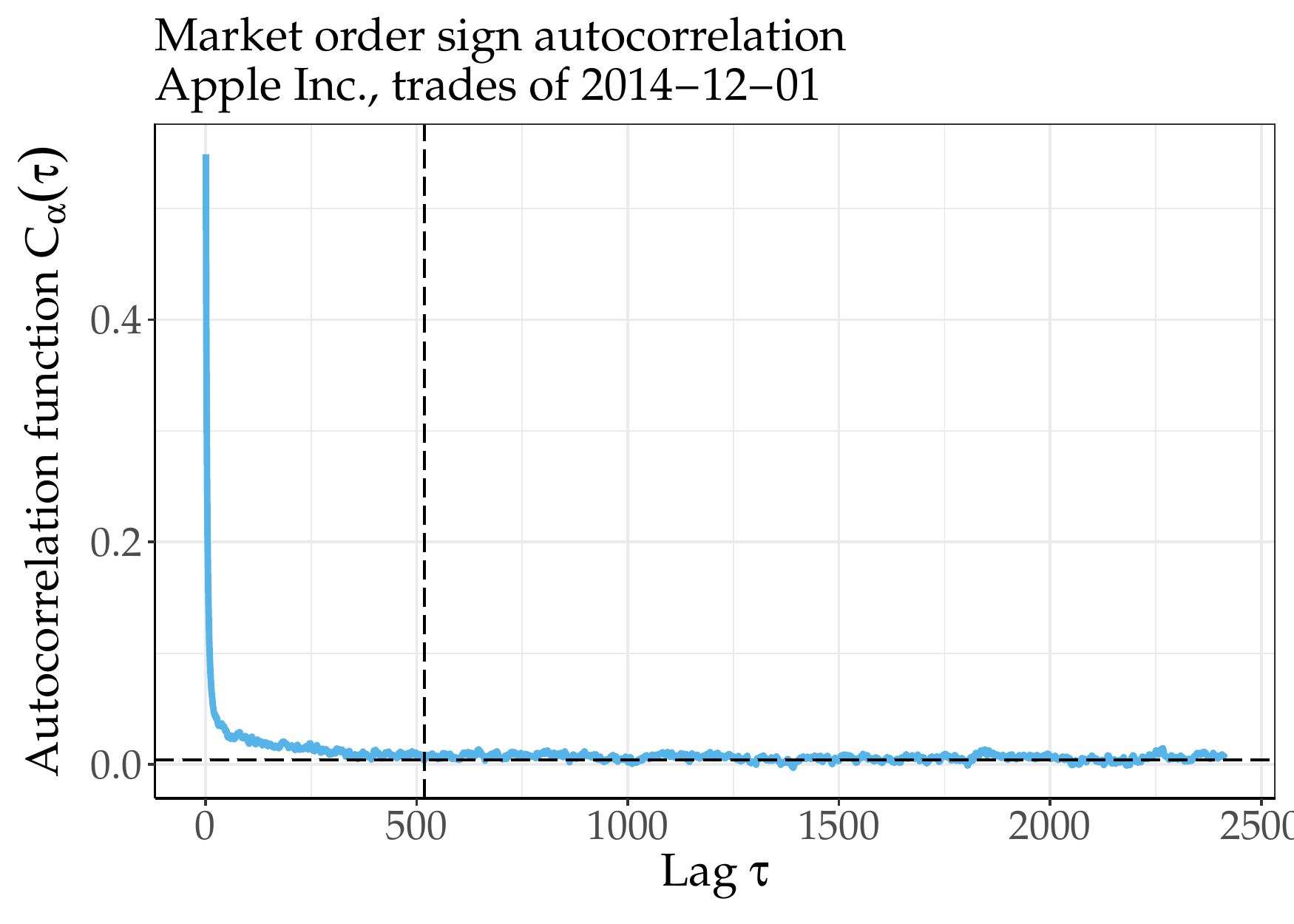}
\par\end{center}%
\end{minipage}\hfill{}%
\begin{minipage}[t]{0.49\columnwidth}%
\begin{center}
\includegraphics[scale=0.4]{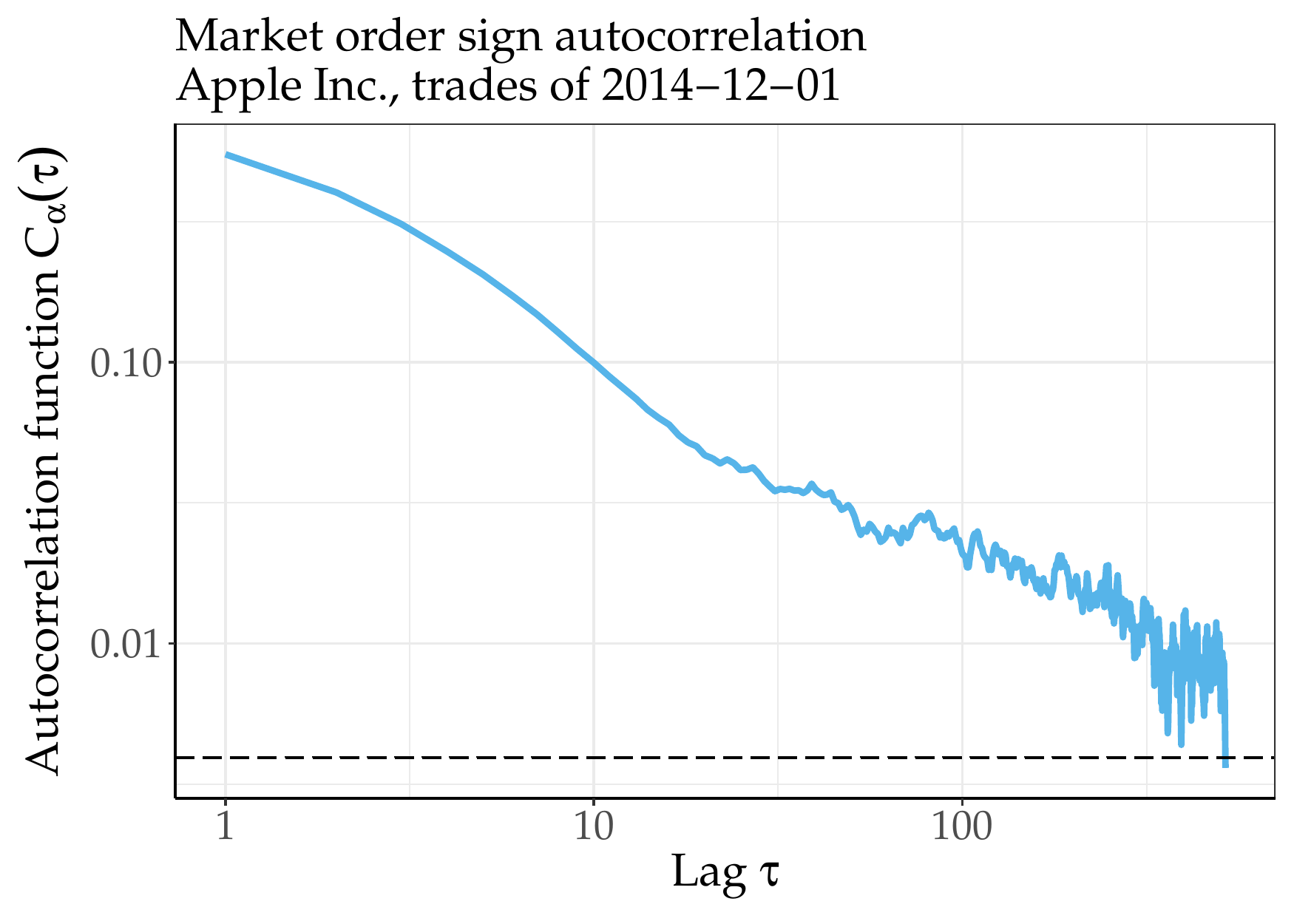}
\par\end{center}%
\end{minipage}
\par\end{centering}
\caption{Example of market order sign autocorrelation functions with linear
axis scales (left plot) and log-log axis scales (right plot). The
black dashed horizontal line is the noise level $2/\sqrt{N}$ where
$N$ is the length of the market order sign time series. The black
dashed vertical line corresponds to the maximum lag $\tau^{*}$. Apple
Inc., trades of 2014-12-01.}
\end{figure}

\subsection{Macro-dynamics: directional fund activity ratio}

First, we introduce a quantity that measures, for a given asset $\alpha$,
the rescaled global directional change of ownership averaged over
all the funds between the quarter ends $q-1$ and $q$. We will call
it the directional fund activity ratio and define it as

\begin{equation}
r_{\alpha}(q)=\frac{\sum_{i}[W_{i\alpha}(q)-W_{i\alpha}(q-1)]}{V_{\alpha}^{{\rm }}(q)},\label{eq:r_alpha}
\end{equation}

where $W_{i\alpha}(q)$ is the position in dollars of fund $i$ on
security $\alpha$ at the end of quarter $q$, and $V_{\alpha}$ the
total volume-dollar of security $\alpha$ exchanged between $q-1$
and $q$. If $r_{\alpha}(t)>0$ (resp. $r_{\alpha}(t)<0$) then the
security $\alpha$ is more bought (resp. sold) than sold (resp. bought)
by the large funds in our database. We will focus on $R_{\alpha}(q)=|r_{\alpha}(q)|$.

\subsection{Macro-dynamics: absolute fund activity ratio}

Another important measure of aggregate fund behaviour consists in
quantifying how much the investment has changed in absolute terms.
We thus define $S_{\alpha}(q)$ as the rescaled absolute difference
of invested amounts between quarter ends $q-1$ and $q$, i.e.,

\begin{equation}
S_{\alpha}(q)=\frac{\sum_{i}|W_{i\alpha}(q)-W_{i\alpha}(q-1)|}{V_{\alpha}(q)}.\label{eq:r_alpha-1}
\end{equation}

This quantity cannot account for round trips of funds over a quarter,
which are fortunately very unlikely for the largest values of $S$
, i.e., the values relevant to the present work. In addition, when
the relative influence on $S$ of large fund round-trips is negligible,
$S$ is a good approximation of large fund participation ratio.

\section{Results}

\subsection{From large fund behaviour to microstructure dynamics}

\begin{figure}
\begin{centering}
\includegraphics[scale=0.4]{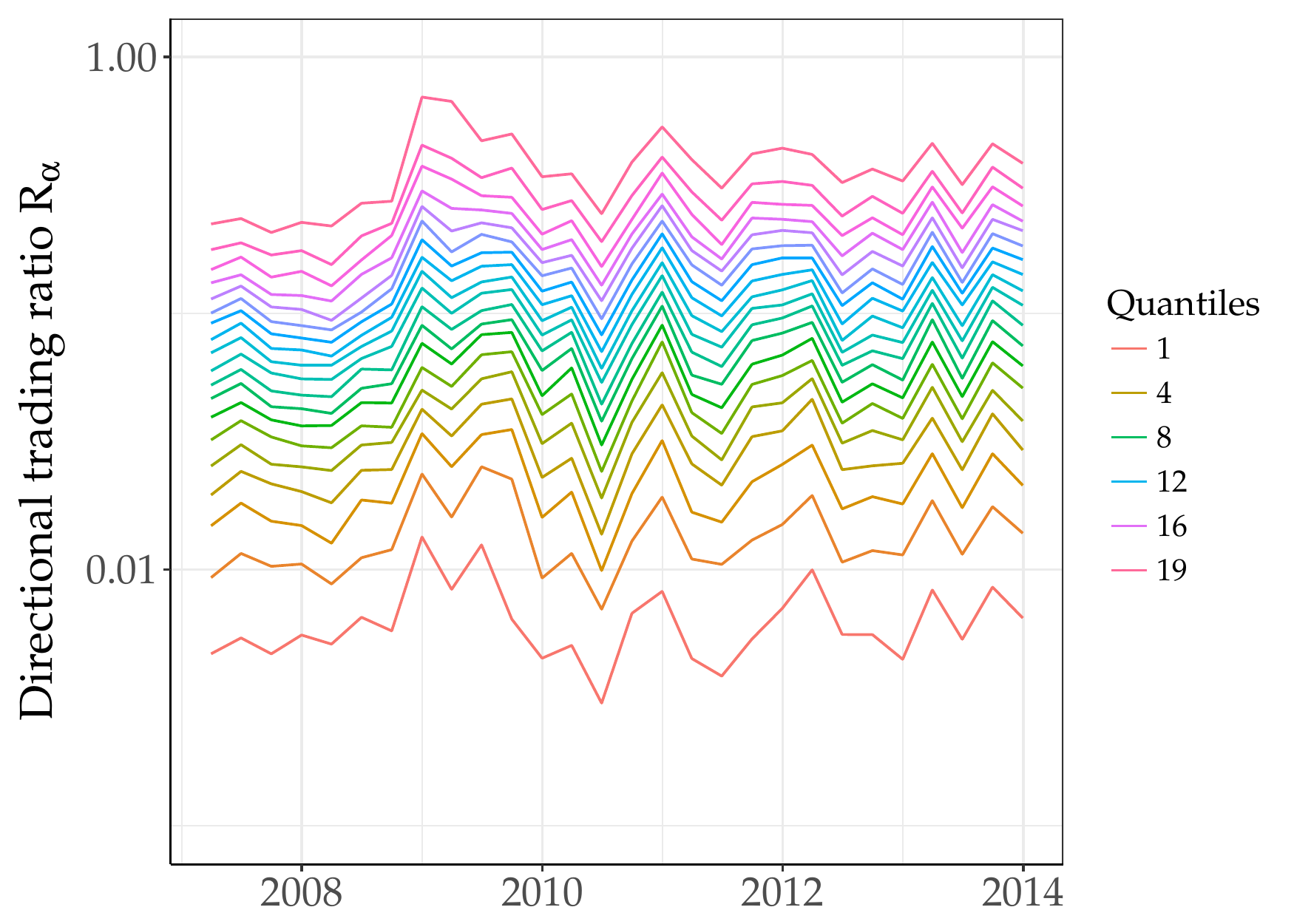} \includegraphics[scale=0.4]{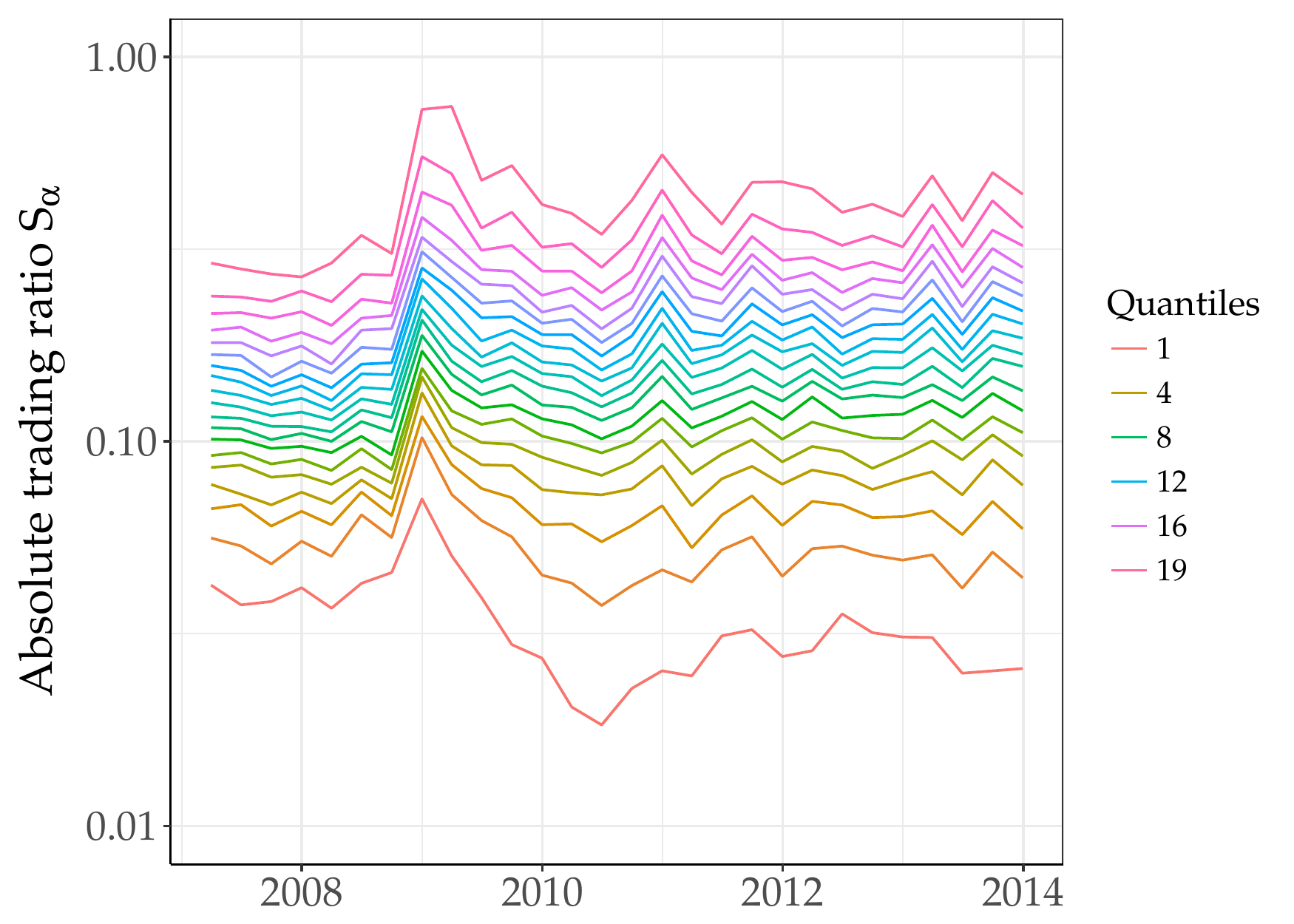}
\par\end{centering}
\caption{Time evolution of the quantiles of the directional fund activity ratio
$R_{\alpha}\left(q\right)$ (left plot) and of the absolute fund activity
ratio $S_{\alpha}\left(q\right)$ (right plot). \label{fig:asset_ownership_variability_t}}
\end{figure}

The premise of this paper is that relating tick-by-tick order book
properties to the fund ownership database is easiest when the aggregate
behaviour of large investment funds is the most extreme, which corresponds
to large values of either $R_{\alpha}$ or $S_{\alpha}$. Thus for
each quarter $q$, we divide the assets into 20 groups of $R_{\alpha}(q)$
by computing the quantiles $k_{R}\in\{1,\cdots,19\}$; we do the same
for $S_{\alpha}(q)$, yielding $k_{S}\in\{1,\cdots,19\}$. We first
compare the microstructural dynamics of the top and bottom groups
of both quantities. By convention, the bottom groups $g_{X}=1$ correspond
to small values of $X\in\{R,S\}$, i.e., to securities that are bought
and sold equally ($R$) or not much traded by large funds ($S$).
Figure \ref{fig:asset_ownership_variability_t} shows the time evolution
of the quantiles of the ratios $R_{\alpha}(q)$ and $S_{\alpha}(q)$.
The large-$R$ quantile are clearly correlated with the large-$S$
quantiles. This should be expected, as a large $R_{\alpha}$ implies
a large $S_{\alpha}$. 

\begin{figure}
\begin{centering}
\begin{minipage}[t]{0.49\columnwidth}%
\begin{center}
\includegraphics[scale=0.4]{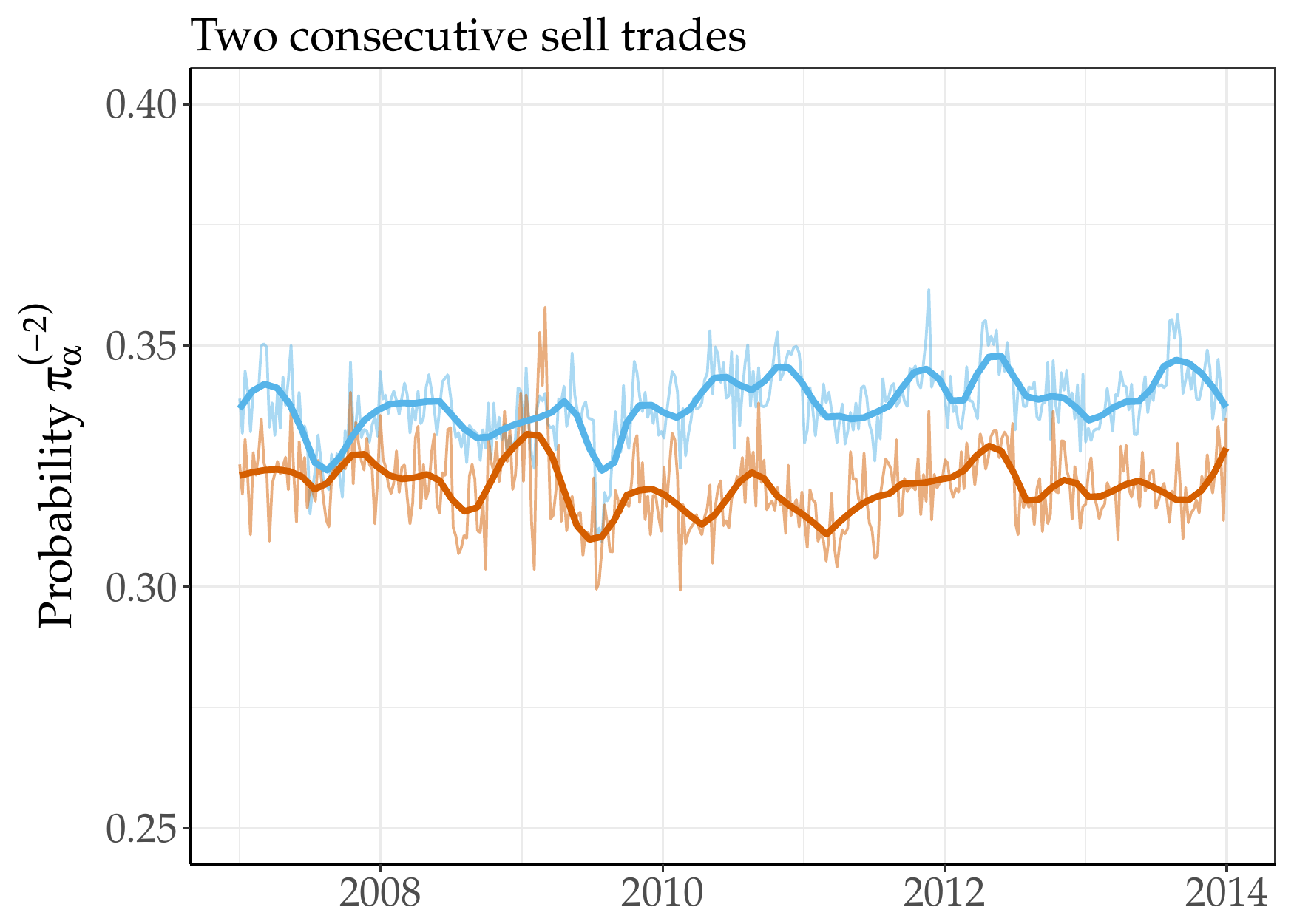}
\par\end{center}%
\end{minipage}%
\begin{minipage}[t]{0.49\columnwidth}%
\begin{center}
\includegraphics[scale=0.4]{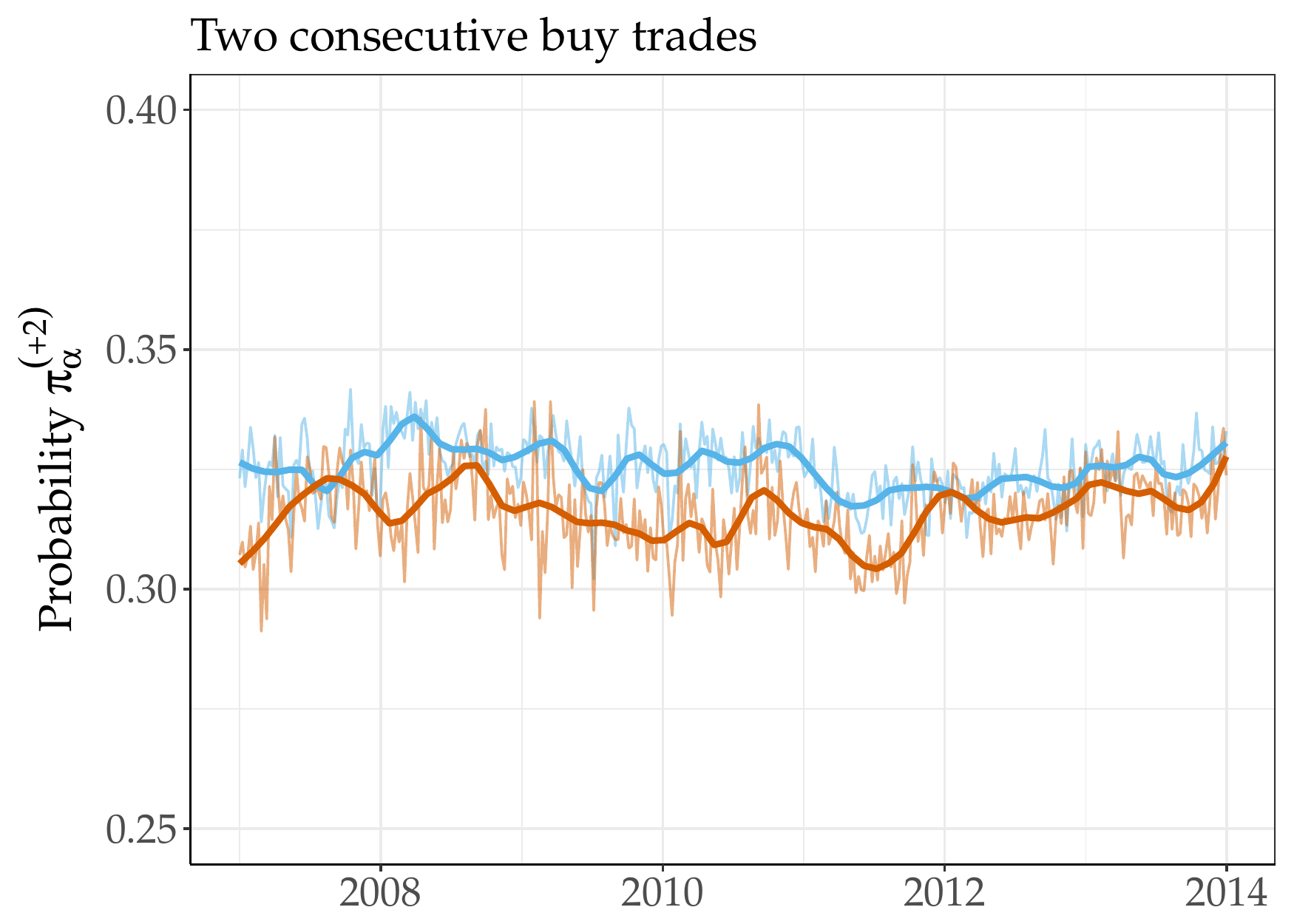}
\par\end{center}%
\end{minipage}
\par\end{centering}
\begin{centering}
\begin{minipage}[t]{0.49\columnwidth}%
\begin{center}
\includegraphics[scale=0.4]{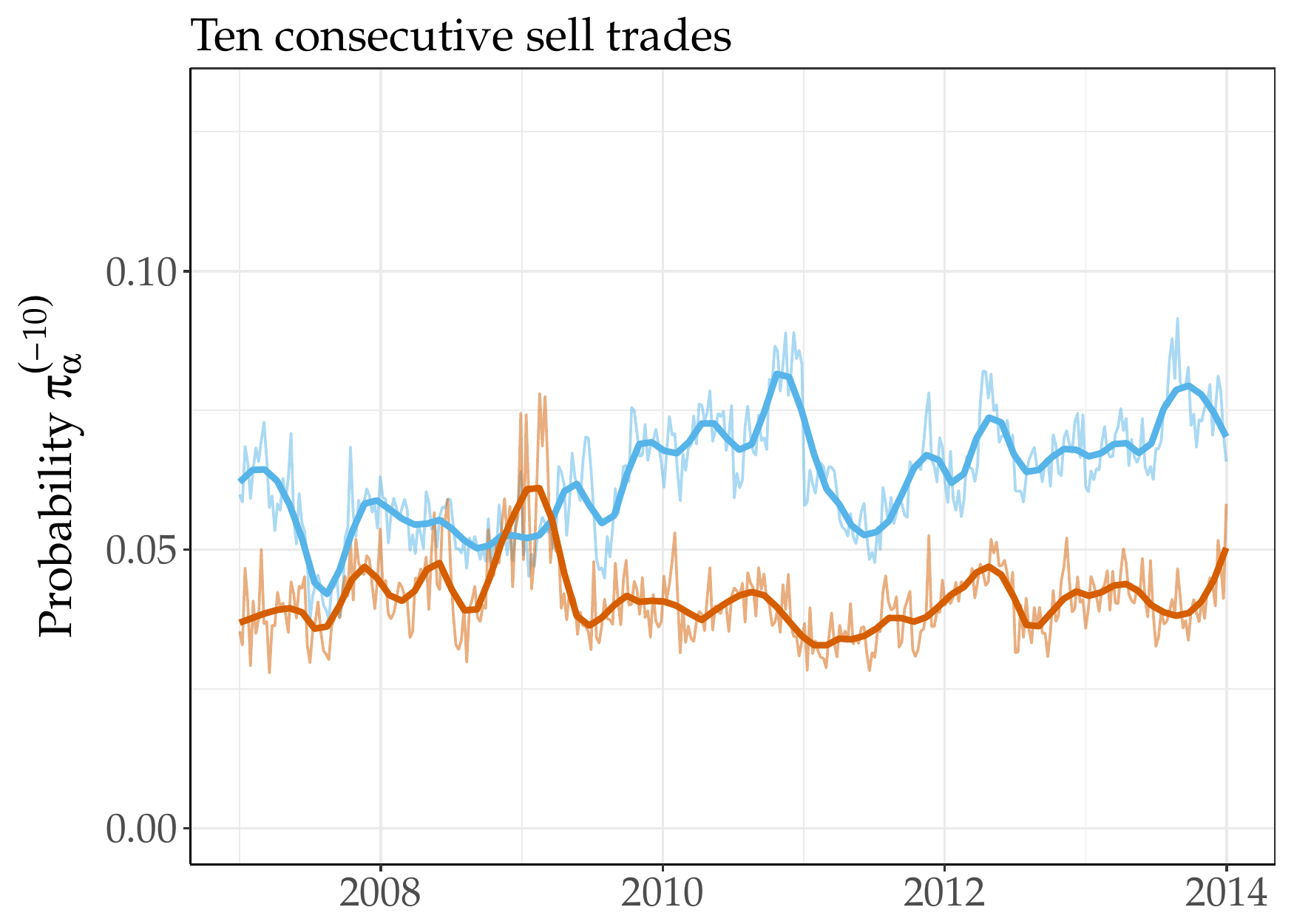}
\par\end{center}%
\end{minipage}%
\begin{minipage}[t]{0.49\columnwidth}%
\begin{center}
\includegraphics[scale=0.4]{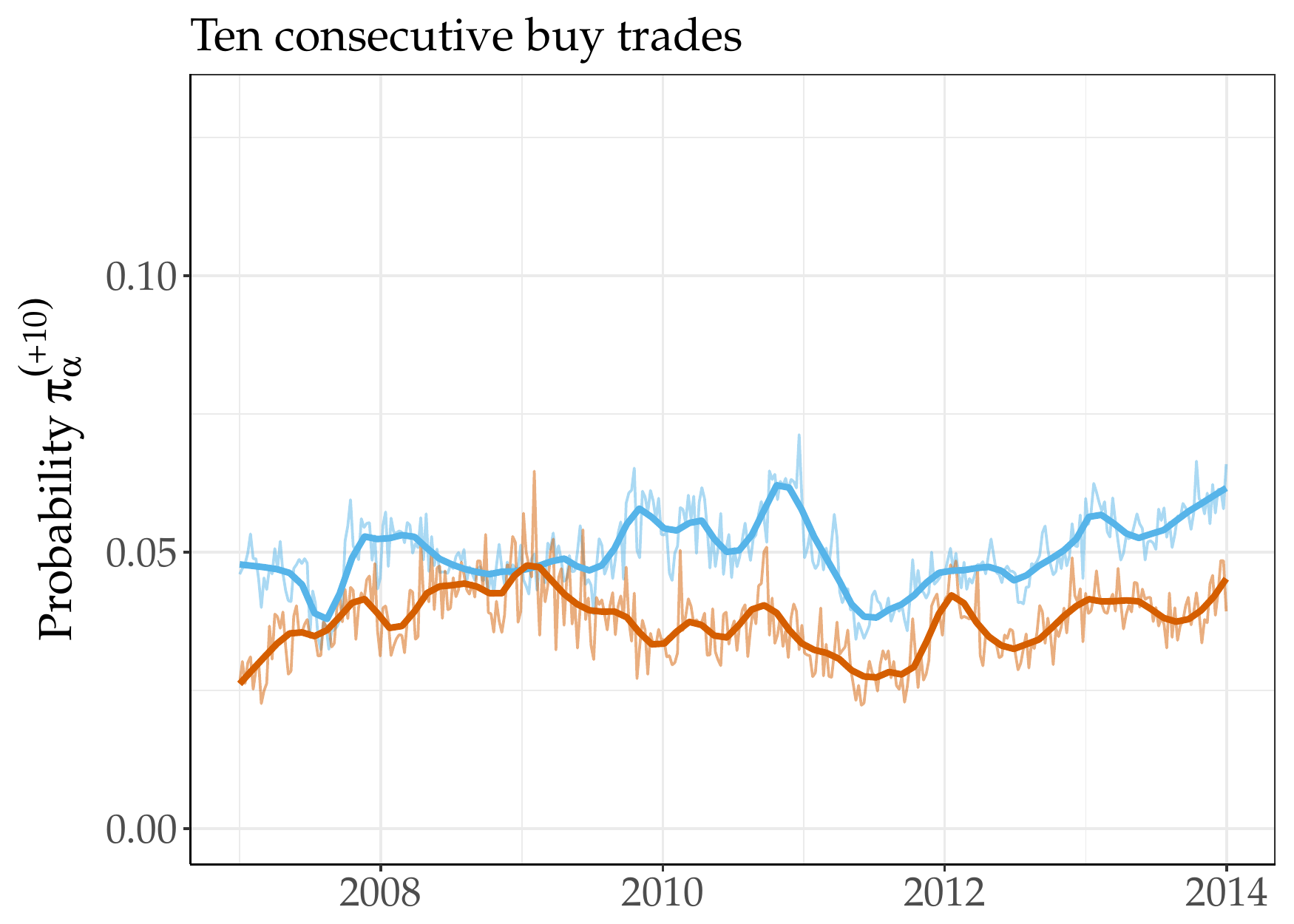}
\par\end{center}%
\end{minipage}
\par\end{centering}
\caption{Time evolution of $\pi_{\alpha}^{(\kappa)}$, the probability of observing
$\kappa$ consecutive negative trade signs (left plots) and $\kappa$
consecutive positive trade signs (right plots) for the top and bottom
quantiles of $R_{\alpha}(q)$ (orange and blue lines, respectively).
\label{fig:prob_2cons_signs-R} }
\end{figure}

\begin{figure}
\begin{centering}
\begin{minipage}[t]{0.49\columnwidth}%
\begin{center}
\includegraphics[scale=0.4]{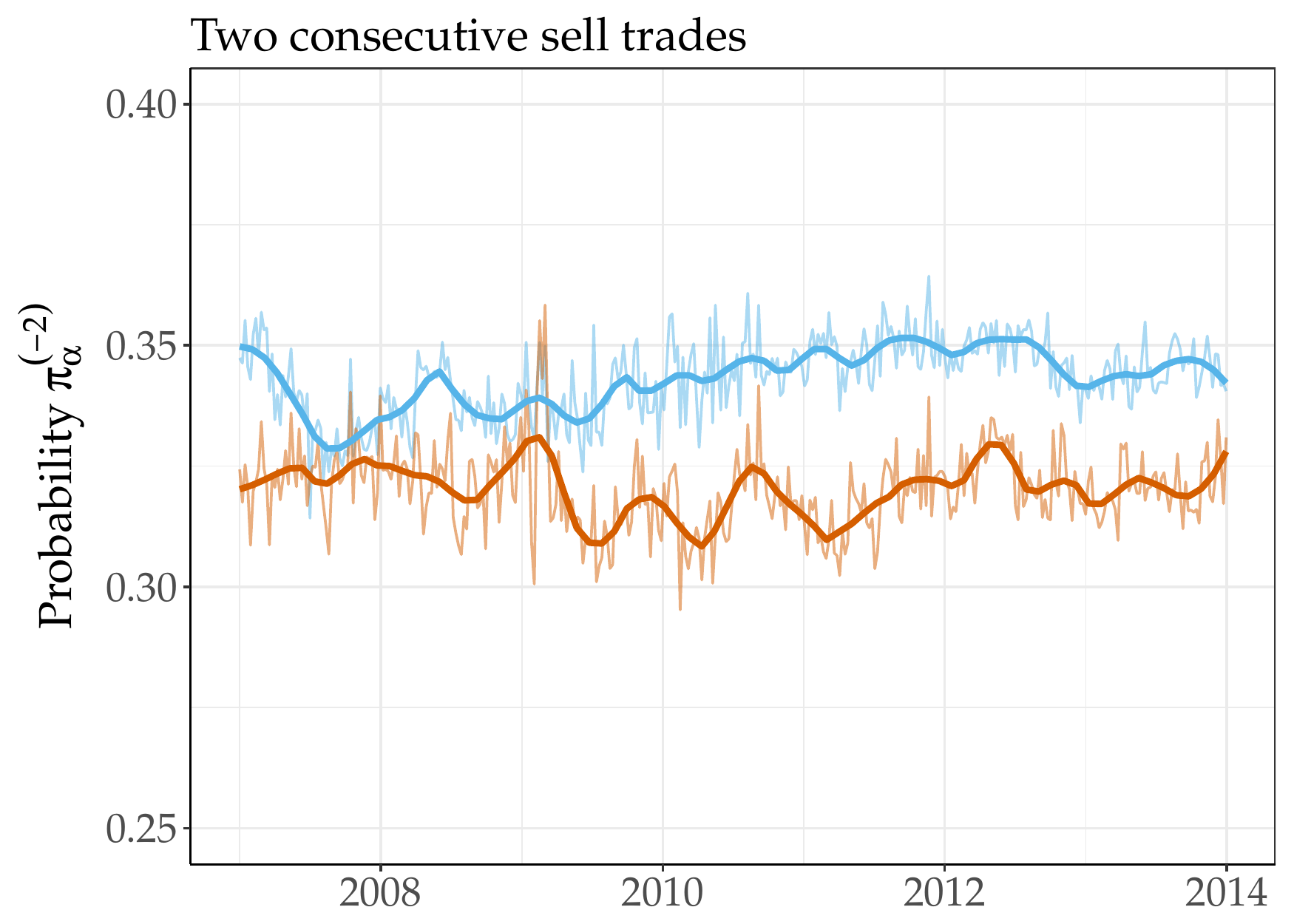}
\par\end{center}%
\end{minipage}\hfill{}%
\begin{minipage}[t]{0.49\columnwidth}%
\begin{center}
\includegraphics[scale=0.4]{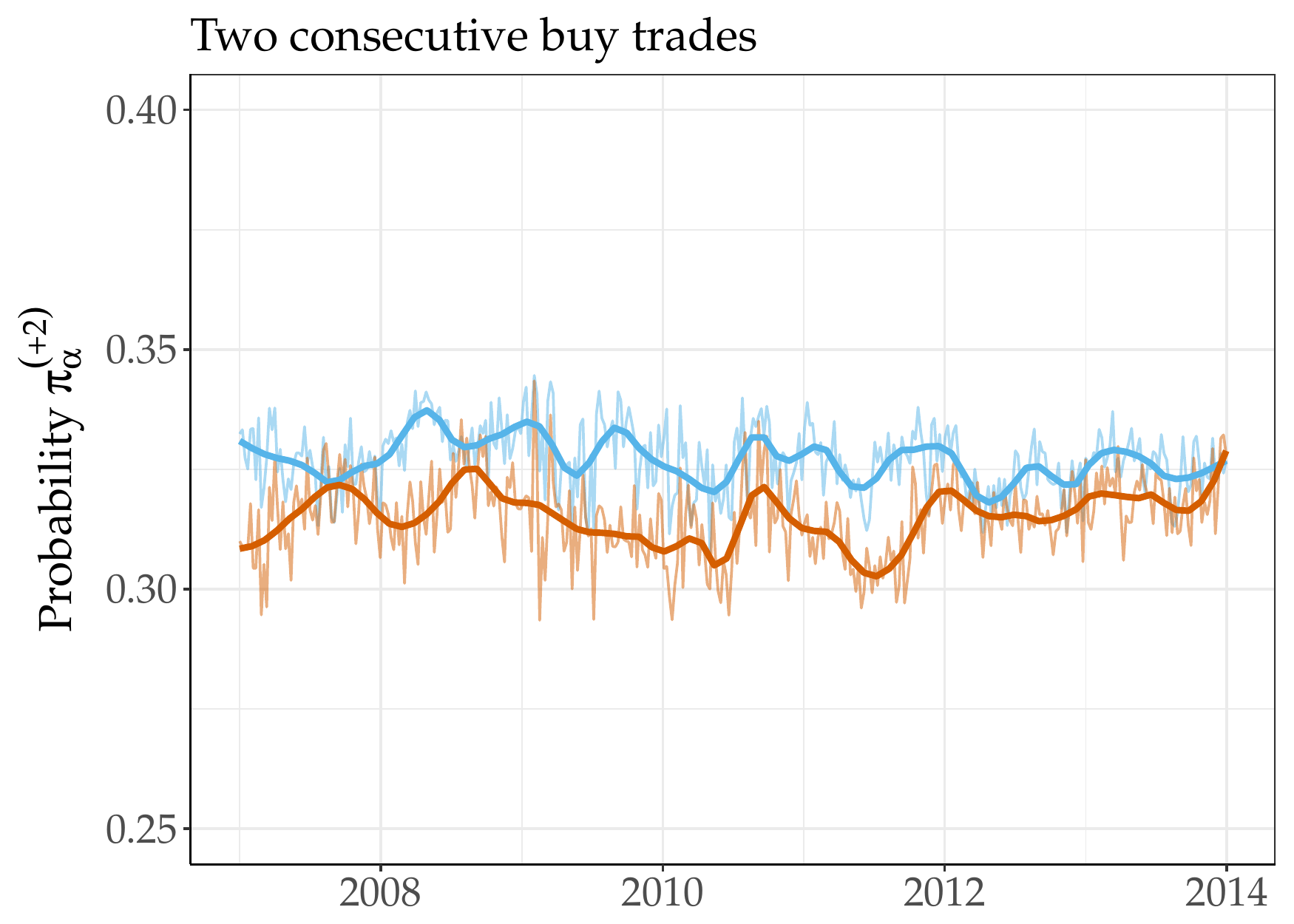}
\par\end{center}%
\end{minipage}
\par\end{centering}
\begin{centering}
\begin{minipage}[t]{0.49\columnwidth}%
\begin{center}
\includegraphics[scale=0.4]{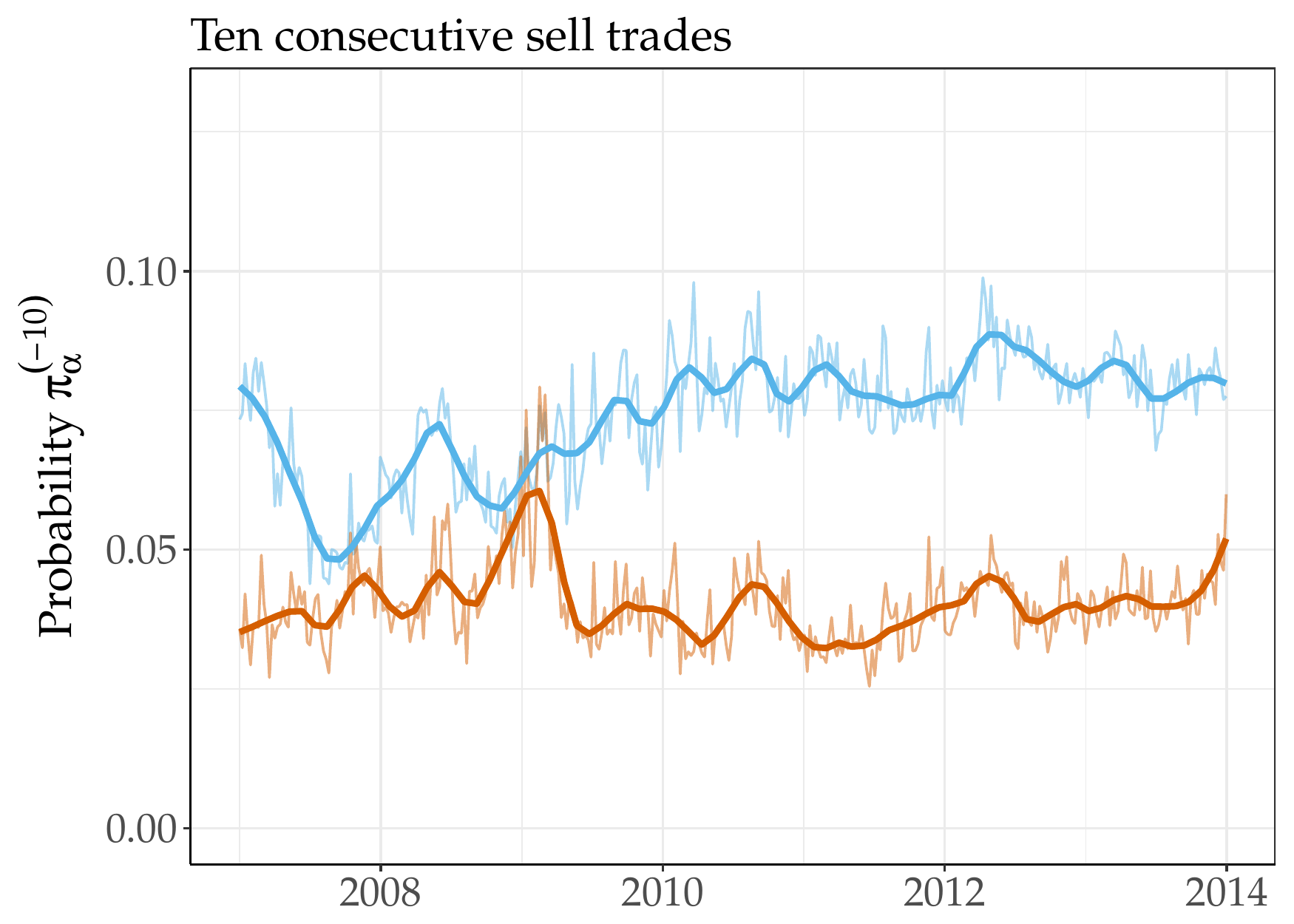}
\par\end{center}%
\end{minipage}\hfill{}%
\begin{minipage}[t]{0.49\columnwidth}%
\begin{center}
\includegraphics[scale=0.4]{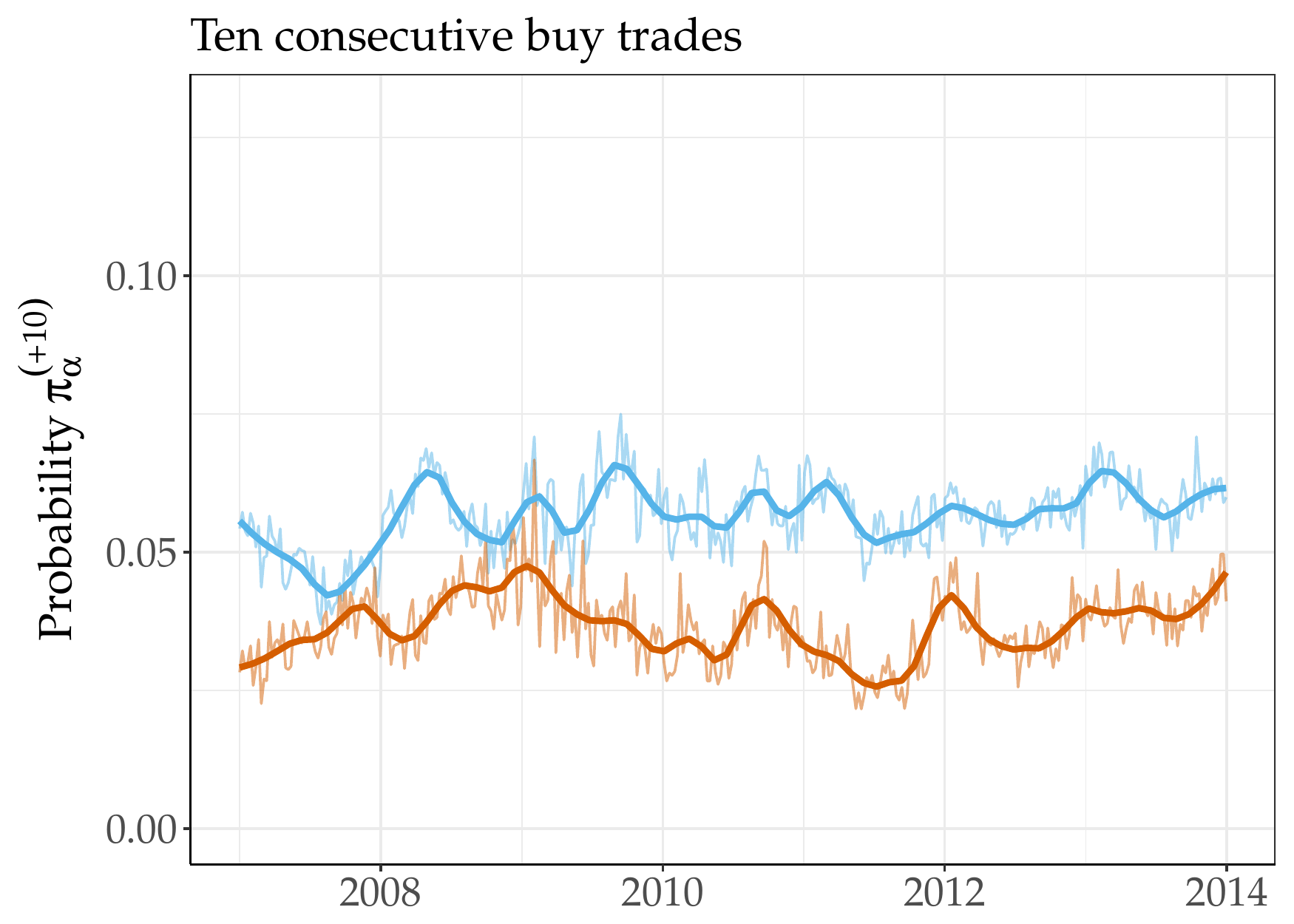}
\par\end{center}%
\end{minipage}
\par\end{centering}
\caption{Time evolution of $\pi_{\alpha}^{(\kappa)}$, the probability of observing
$\kappa$ consecutive negative trade signs (left plots) and $\kappa$
consecutive positive trade signs (right plots) for the top and bottom
quantiles of $S_{\alpha}(q)$ (orange and blue lines, respectively).
\label{fig:prob_2cons_signs-S} }
\end{figure}

Focusing on the assets belonging to the top and bottom groups determined
by the quantiles of $R$ and $S$, one now assess the influence of
trading by large funds on the market order sign memory length measures.
Starting with the two extreme groups determined by the quantiles of
$R$, Fig\@.~\ref{fig:prob_2cons_signs-R} reports that the frequency
of $\kappa$ consecutive trades of the same sign $\pi_{\alpha}^{(s\kappa)}$
for $\kappa=2$, once averaged over all assets belonging to a given
group, is consistently different between the two groups of assets
as time goes on; one also sees that the difference is larger for $\kappa=10$
than for $\kappa=2$; in fact, it is an increasing function of $\kappa$,
at least for $2\le\kappa\le10$. The difference is larger when the
assets are grouped according to the quantiles of $S$, as illustrated
in Fig\@.\ref{fig:prob_2cons_signs-S}. In short, being actively
traded by large funds decreases the probability of occurrence of consecutive
market orders of the same kind, which thus is a sign of weakening
of market order sign memory.

\begin{figure}
\begin{centering}
\begin{minipage}[t]{0.49\columnwidth}%
\begin{center}
\includegraphics[scale=0.4]{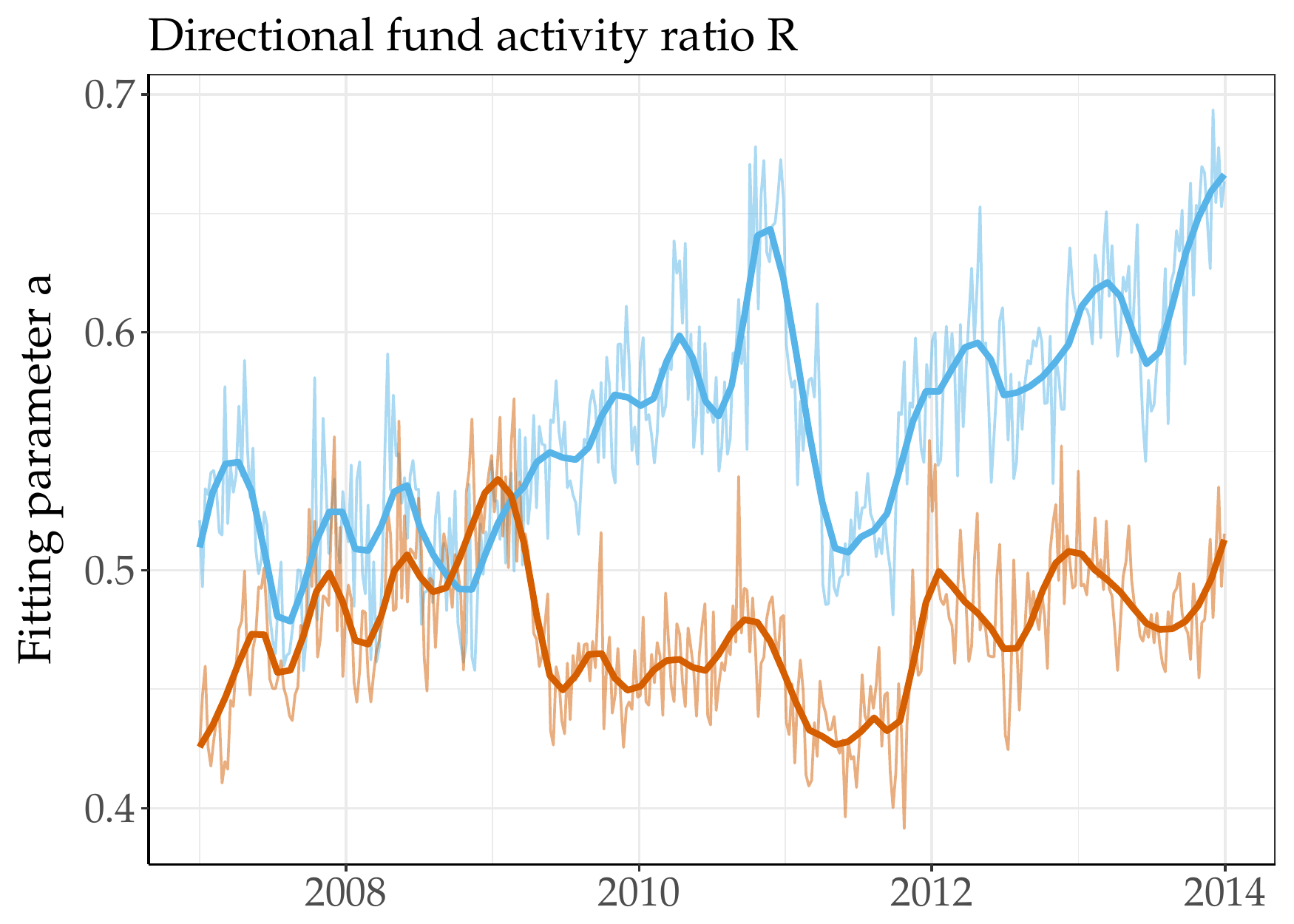}
\par\end{center}%
\end{minipage}\hfill{}%
\begin{minipage}[t]{0.49\columnwidth}%
\begin{center}
\includegraphics[scale=0.4]{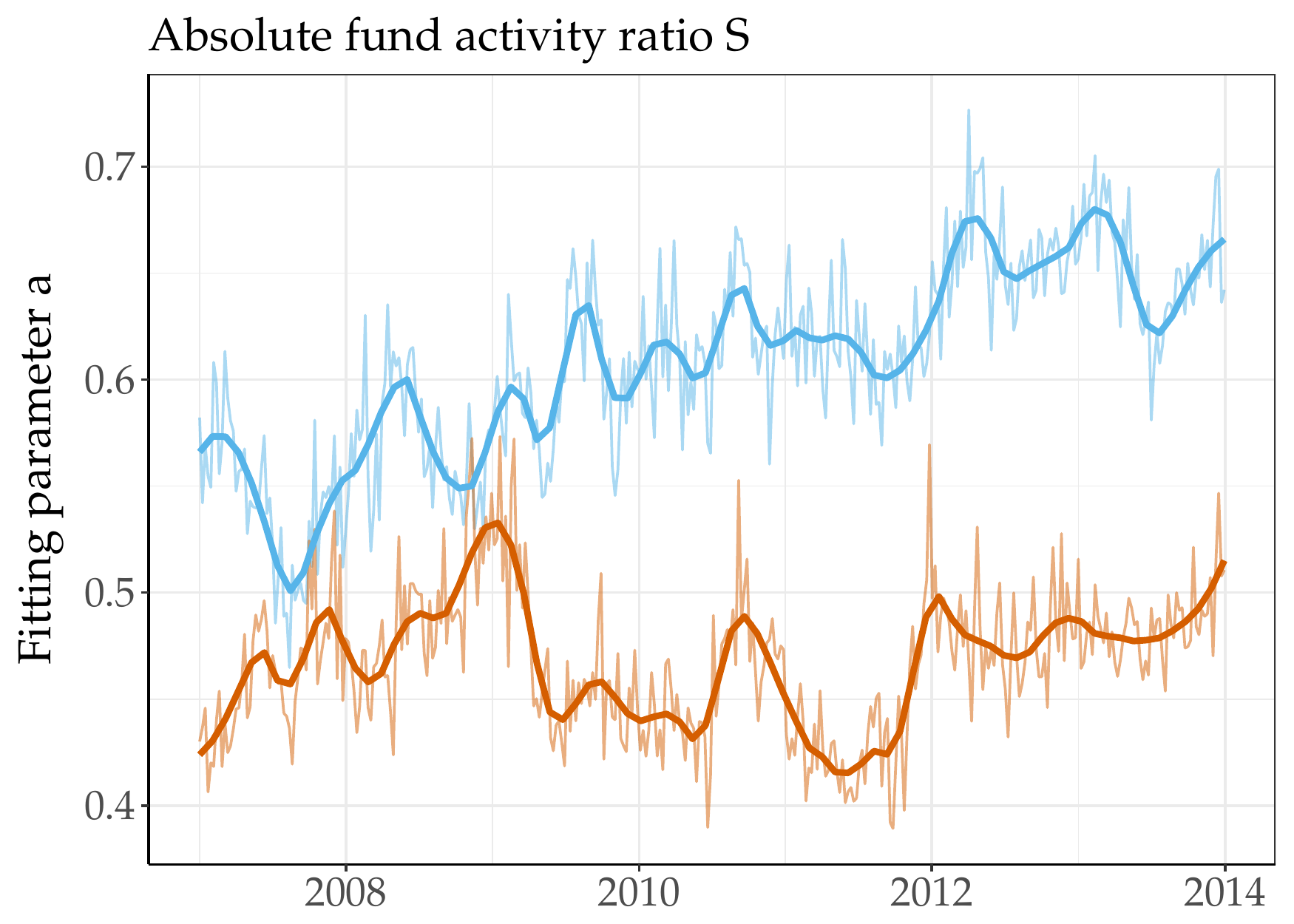}
\par\end{center}%
\end{minipage}
\par\end{centering}
\begin{centering}
\begin{minipage}[t]{0.49\columnwidth}%
\begin{center}
\includegraphics[scale=0.4]{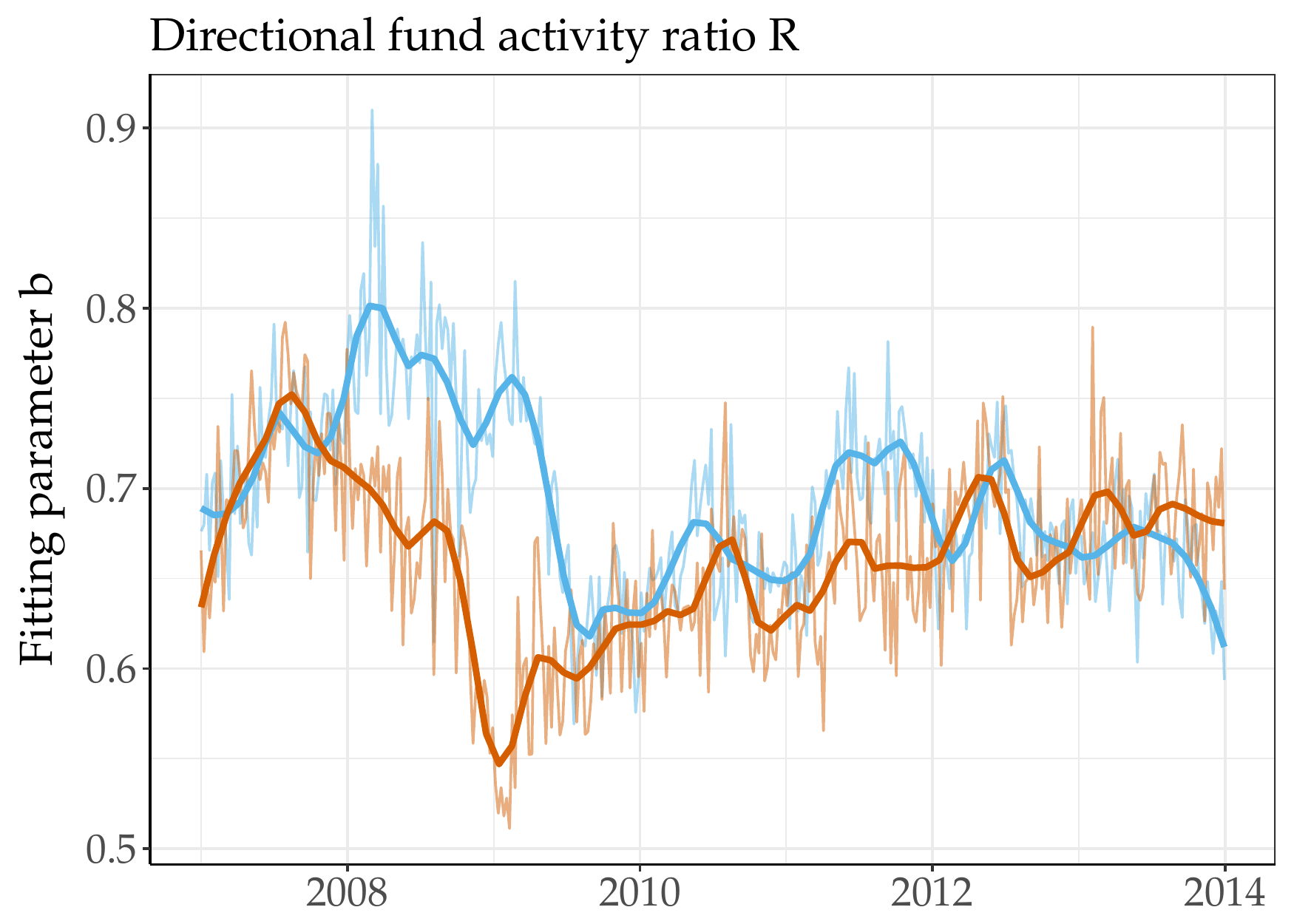}
\par\end{center}%
\end{minipage}\hfill{}%
\begin{minipage}[t]{0.49\columnwidth}%
\begin{center}
\includegraphics[scale=0.4]{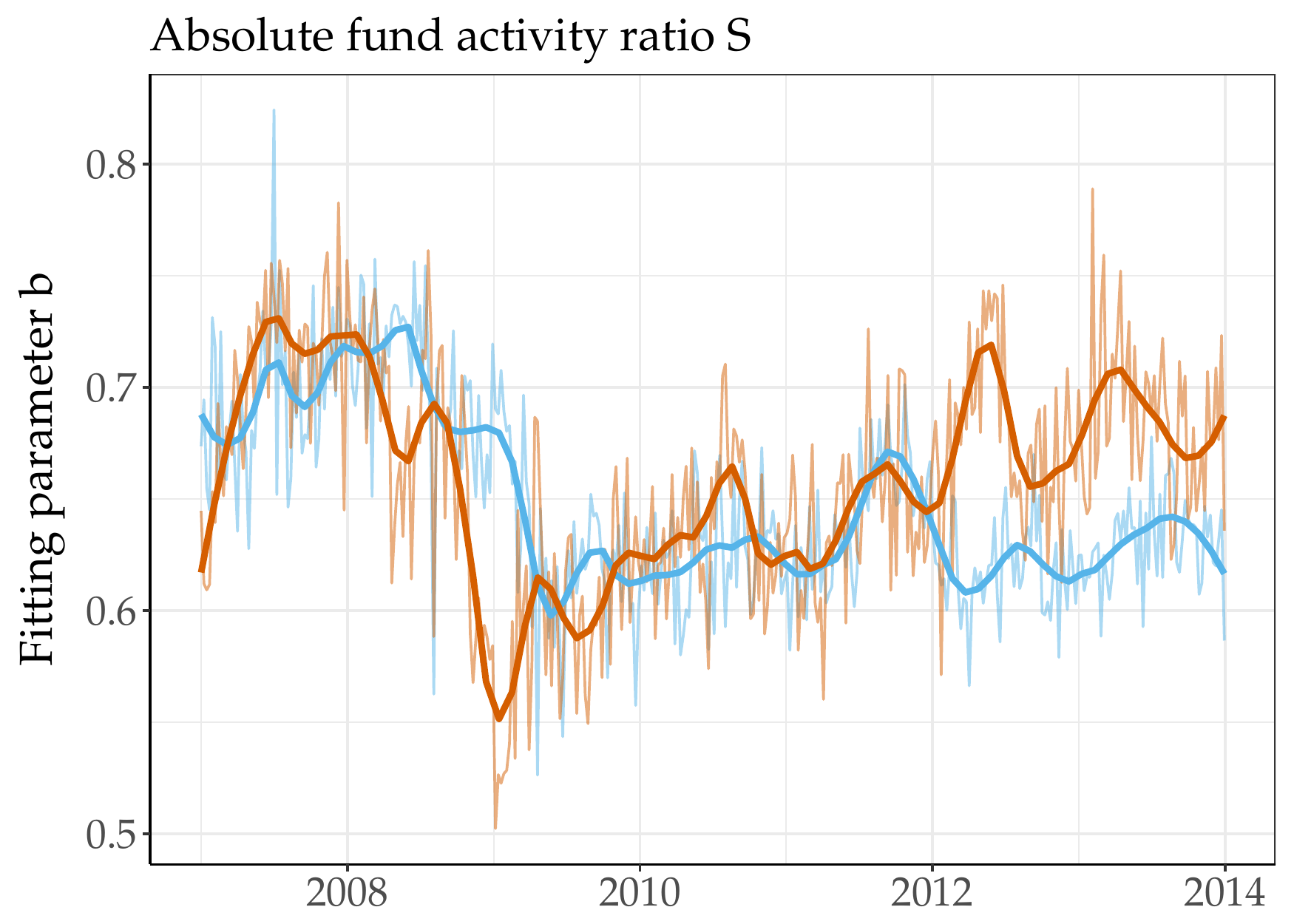}
\par\end{center}%
\end{minipage}
\par\end{centering}
\caption{Time evolution of the fitting parameters characterizing the long memory
of the market order sign autocorrelation $C_{\alpha}(\tau)=a\tau^{b}$.
Top plots: evolution of $a$ averaged over all the members of the
bottom and top quantile of $R_{\alpha}(q)$ (left plot) and $S_{\alpha}\left(q\right)$
(right plot). Bottom plots: evolution of $b$ averaged over all the
members of the bottom and top quantiles (orange and blue lines, respectively)
of $R_{\alpha}(q)$ (left plot) and $S_{\alpha}(q)$ (right plot).
\label{fig:auto_exps} }
\end{figure}

The other measures of memory length lead to the same conclusion. For
example, fitting the trade sign autocorrelation $C_{\alpha}(\tau)$
with $a\tau^{-b}$ for $\alpha$ in the top and bottom groups of assets
consistently yields smaller values of the prefactor $a$ for the quantiles
of either $R$ or $S$ (top plots of Fig.~\ref{fig:auto_exps}),
except in 2008-2009 with respect to the quantiles of $R$: during
this period, $a$ was roughly the same in both groups. The similar
behaviour of $\pi_{\alpha}^{(s\kappa)}$ and $a$ is to be expected:
$C_{\alpha}(\tau)$ being a function of $\{\pi_{\alpha}^{(s\kappa)}\}_{\kappa}$,
the prefactor $a$ is mostly related to small-$\kappa$ probabilities
$\pi_{\alpha}^{(s\kappa)}$. The case of the exponent $b$ is more
nuanced and revealingly so (bottom-left plot of Fig.~\ref{fig:auto_exps}):
large directional trading by large funds has no clear influence on
$b$, except in times of crisis, as e.g. in 2008-2009 when assets
with large $R$ a smaller $b$ than the assets in the small-$R$ group.
The 2011 crisis also lead to a significant and similar influence of
$R$ on $b$; while the typical value of $a$ of assets with a large
$R$ plunged, it did not reach that of assets with small $R$, contrarily
to what happened during the 2008-2009 period. The absolute activity
ratio $S$ has always been discriminant for $a$. Regarding $b$,
assets with a large $S$ also had a smaller $b$ (but a large $a$)
in 2008-2009, while in the 2012-2014 period, the reverse is true.
Thus these fitting parameters provide a more dynamic picture on the
influence of the activity of large funds.

\begin{figure}
\begin{centering}
\begin{minipage}[t]{0.49\columnwidth}%
\begin{center}
\includegraphics[scale=0.4]{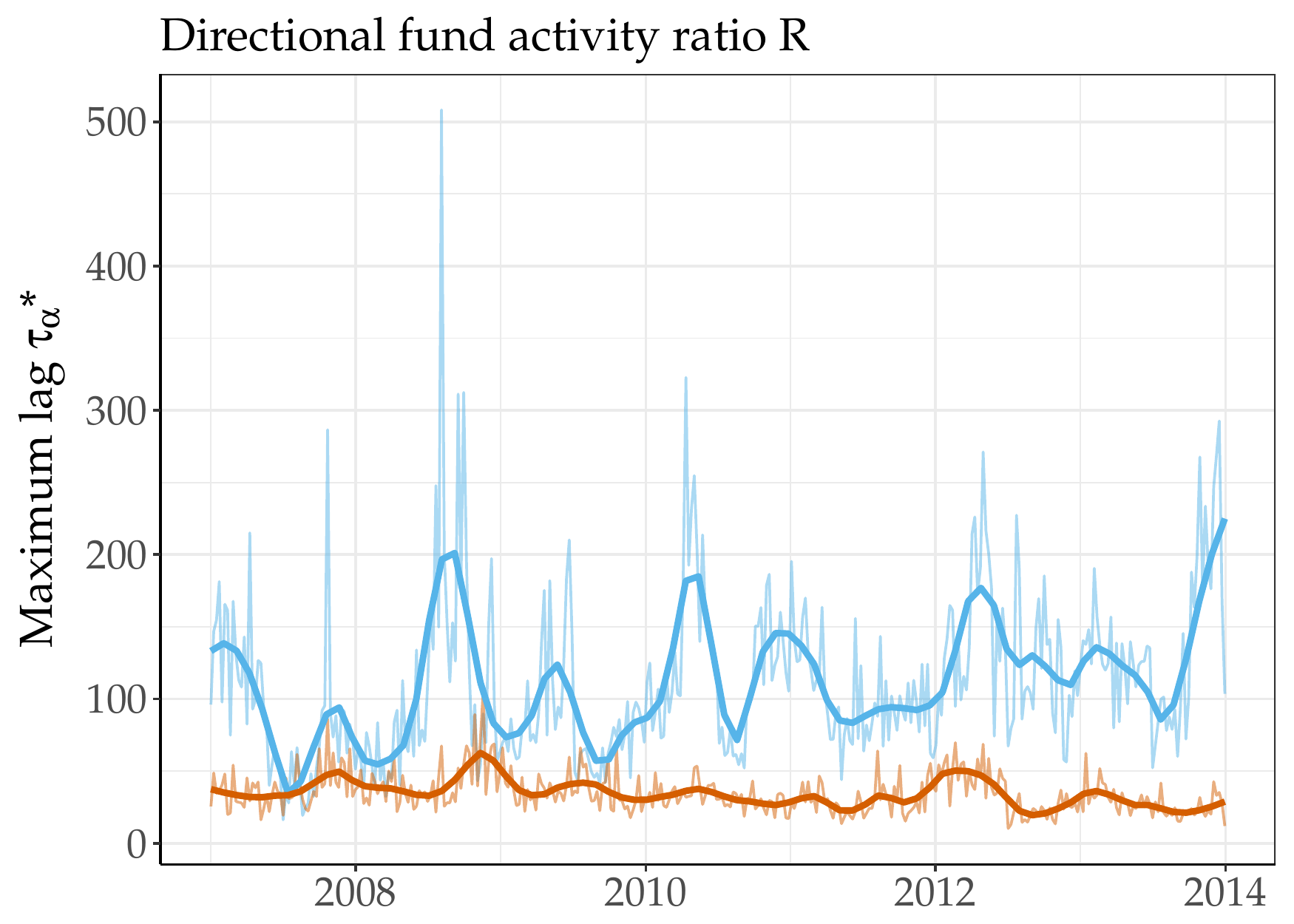}
\par\end{center}%
\end{minipage}\hfill{}%
\begin{minipage}[t]{0.49\columnwidth}%
\begin{center}
\includegraphics[scale=0.4]{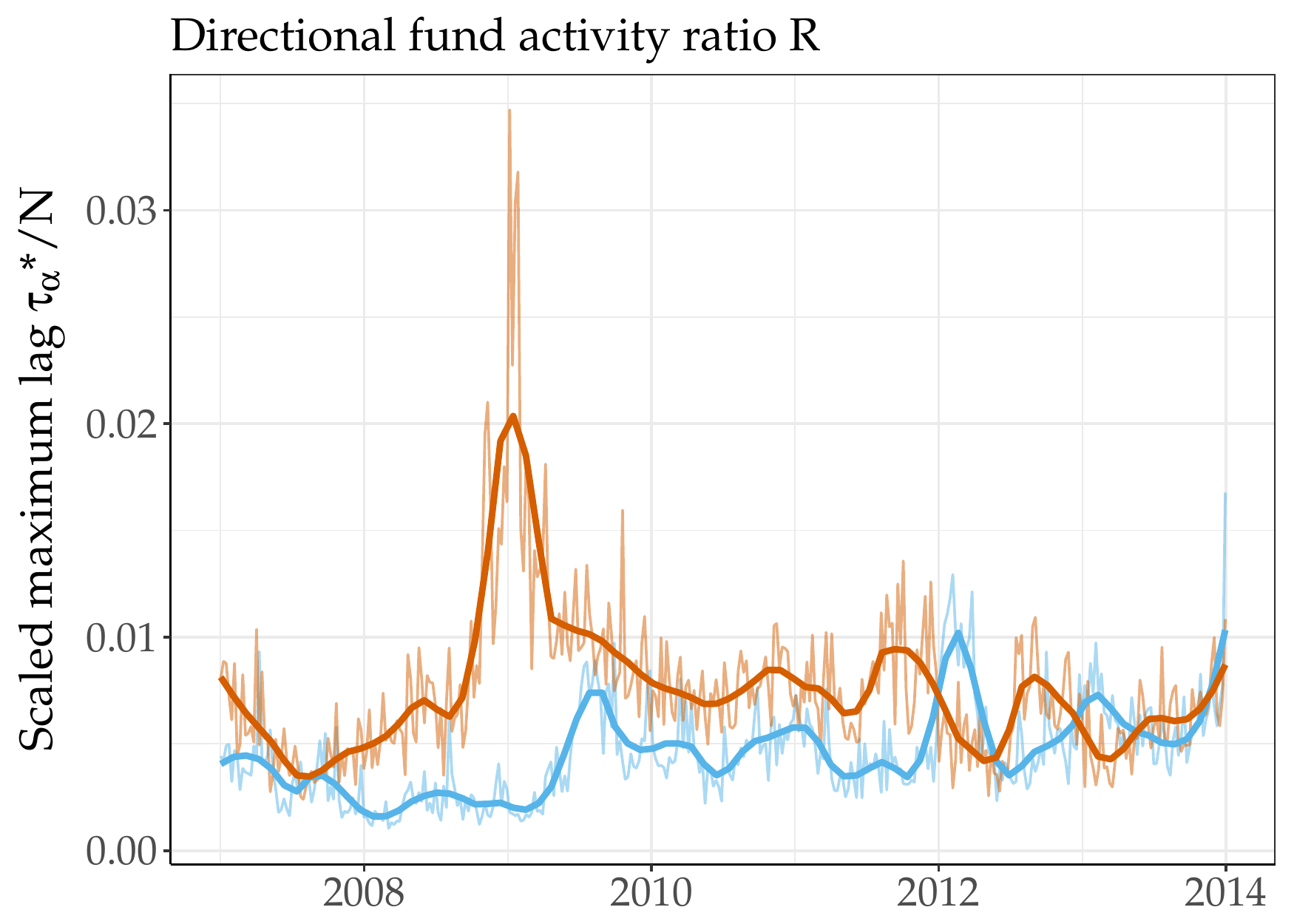}
\par\end{center}%
\end{minipage}
\par\end{centering}
\begin{centering}
\begin{minipage}[t]{0.49\columnwidth}%
\begin{center}
\includegraphics[scale=0.4]{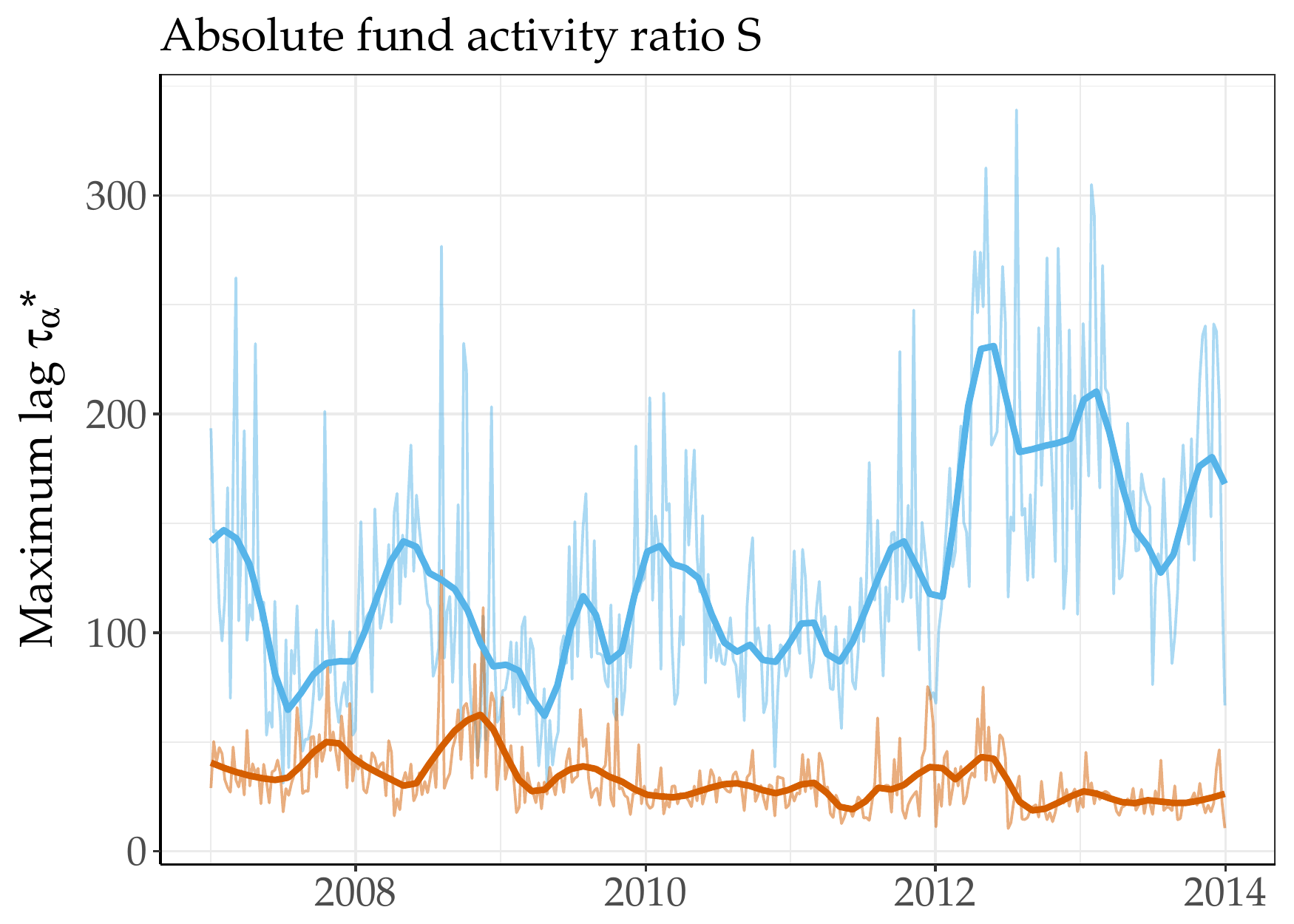}
\par\end{center}%
\end{minipage}\hfill{}%
\begin{minipage}[t]{0.49\columnwidth}%
\begin{center}
\includegraphics[scale=0.4]{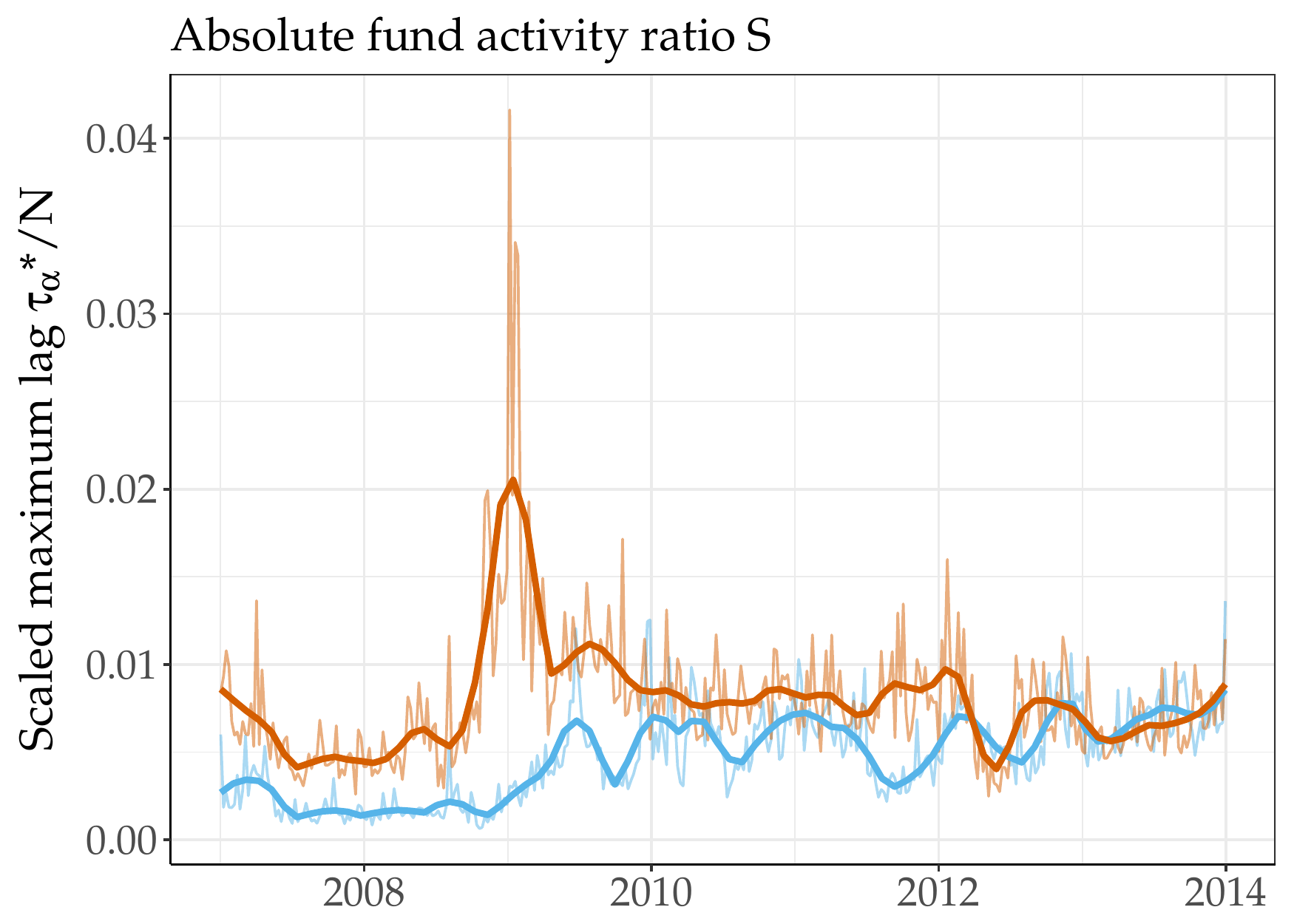}
\par\end{center}%
\end{minipage}
\par\end{centering}
\caption{Time evolution of $\tau_{\alpha}^{*}$, the scaled maximum lag before
the autocorrelation function of the trade signs reaches the noise
level, averaged over the top and bottom quantiles (orange and blue
lines, respectively) of $R_{\alpha}(q)$ (upper plots) and $S_{\alpha}(q)$
(bottom plots).\label{fig:tau}}
\end{figure}

The overall picture is nevertheless the same: for a given number of
trades, on average, the difference of $a$ and $b$ between the top
and bottom quantiles of $R$ and $S$ contribute to shorten the length
of the memory as inferred by $C_{\alpha}(\tau)$ because the noise
level is reached at smaller lags for assets in the top quantiles.
This is indeed confirmed by Fig.~\ref{fig:tau} which shows the time
evolution of the $\tau_{\alpha}^{*}$ averaged over the top and bottom
quantiles of $R$ and $S$. One notes that $\tau_{\alpha}^{*}$ of
the top and bottom groups are clearly separated, while this ceases
to be the case for and $\tau_{\alpha}^{*}/N$ since 2012. The effect
of $R$ or $S$ is opposite on $\tau_{\alpha}^{*}$ and $\tau_{\alpha}^{*}/N$,
which is due to the fact that the number of transactions $N$ of assets
with large $R$ or $S$ is typically smaller.


\subsection{Large fund directional and absolute trading detection}

A more relevant question in practice is whether one can detect trading
by large investment funds from quantities measured from tick-by-tick
data. In the context of this paper, the question may be rephrased
as how to guess in which quantile of $R$ or $S$ a given asset may
be from the knowledge of $a$, $b$, $\tau^{*}$ or $\tau^{*}/N$.
Since the later quantities are measured of a week, we compute with
their averages over a given quarter.

\begin{figure}
\begin{minipage}[t]{0.49\columnwidth}%
\begin{center}
\includegraphics[scale=0.4]{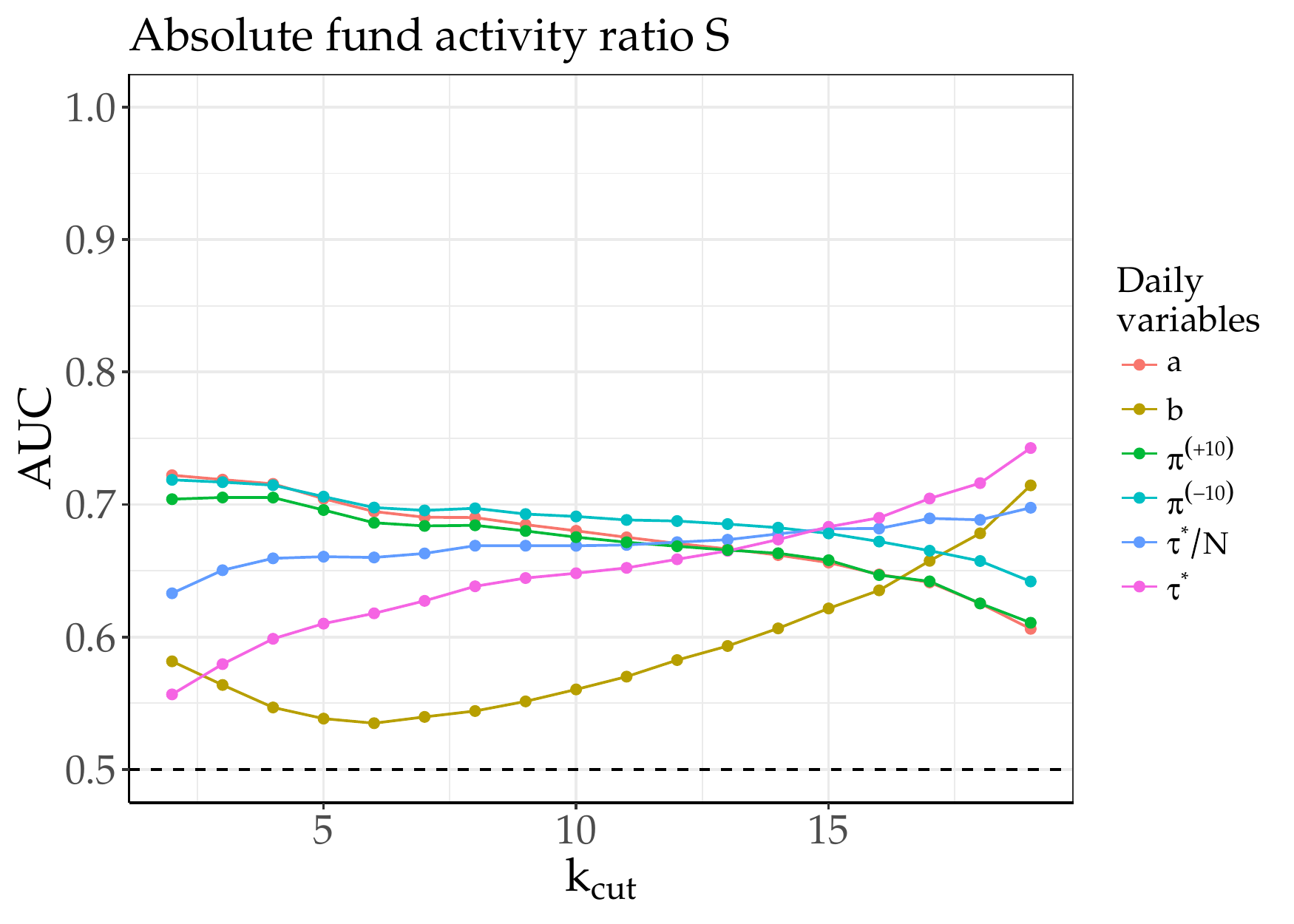}
\par\end{center}%
\end{minipage}\hfill{}%
\begin{minipage}[t]{0.49\columnwidth}%
\begin{center}
\includegraphics[scale=0.4]{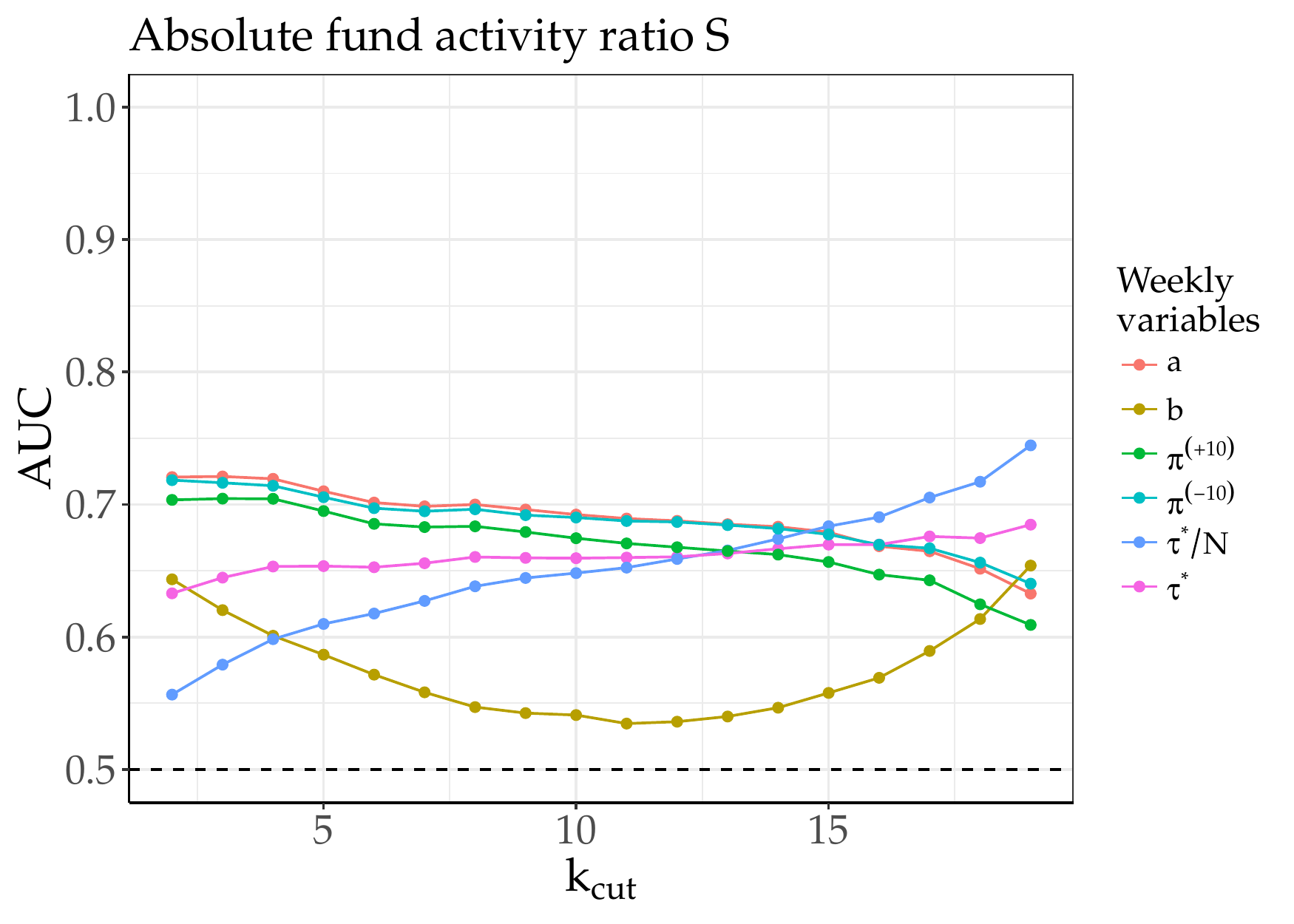}
\par\end{center}%
\end{minipage}
\begin{centering}
\begin{minipage}[t]{0.49\columnwidth}%
\begin{center}
\includegraphics[scale=0.4]{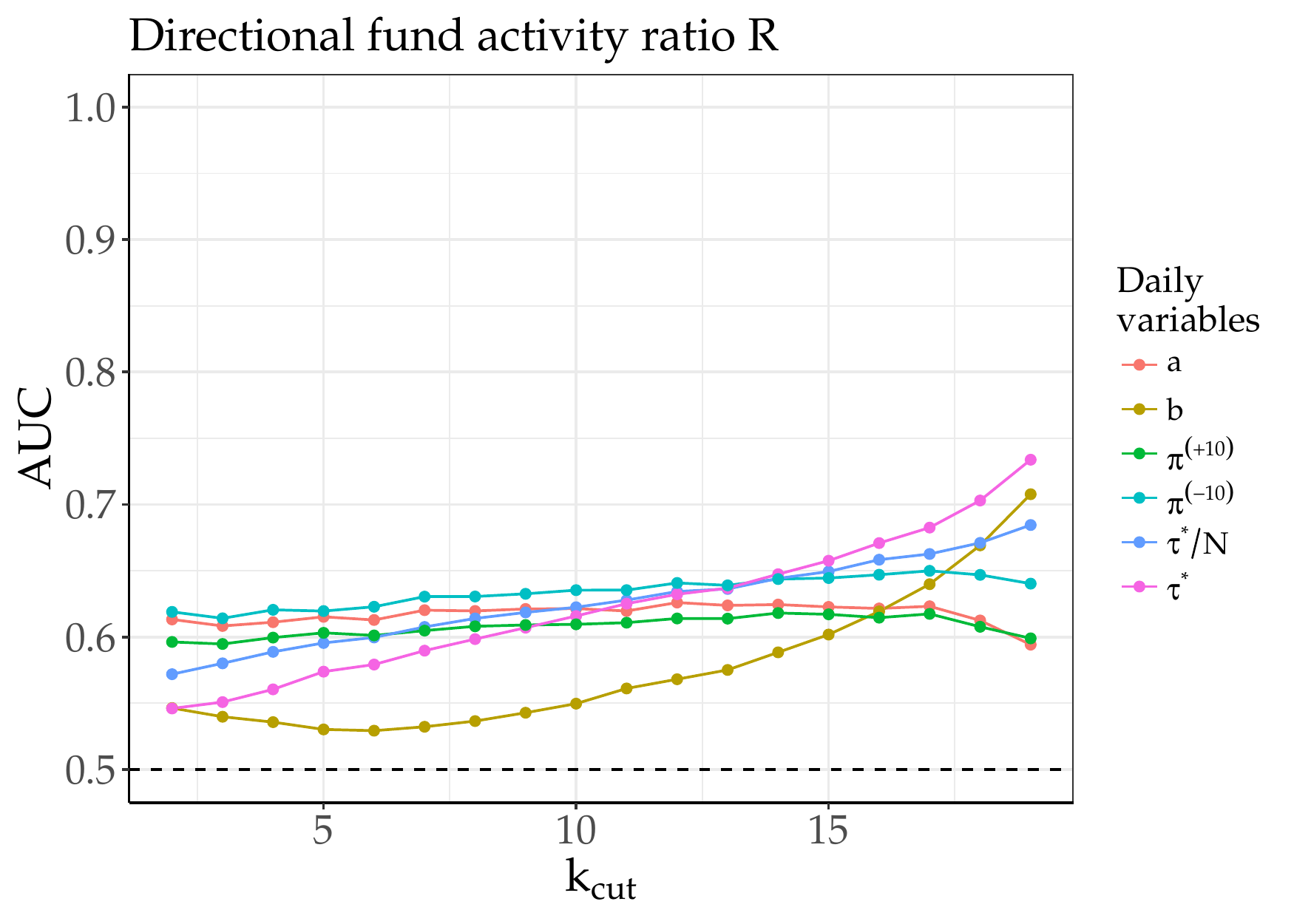}
\par\end{center}%
\end{minipage}\hfill{}%
\begin{minipage}[t]{0.49\columnwidth}%
\begin{center}
\includegraphics[scale=0.4]{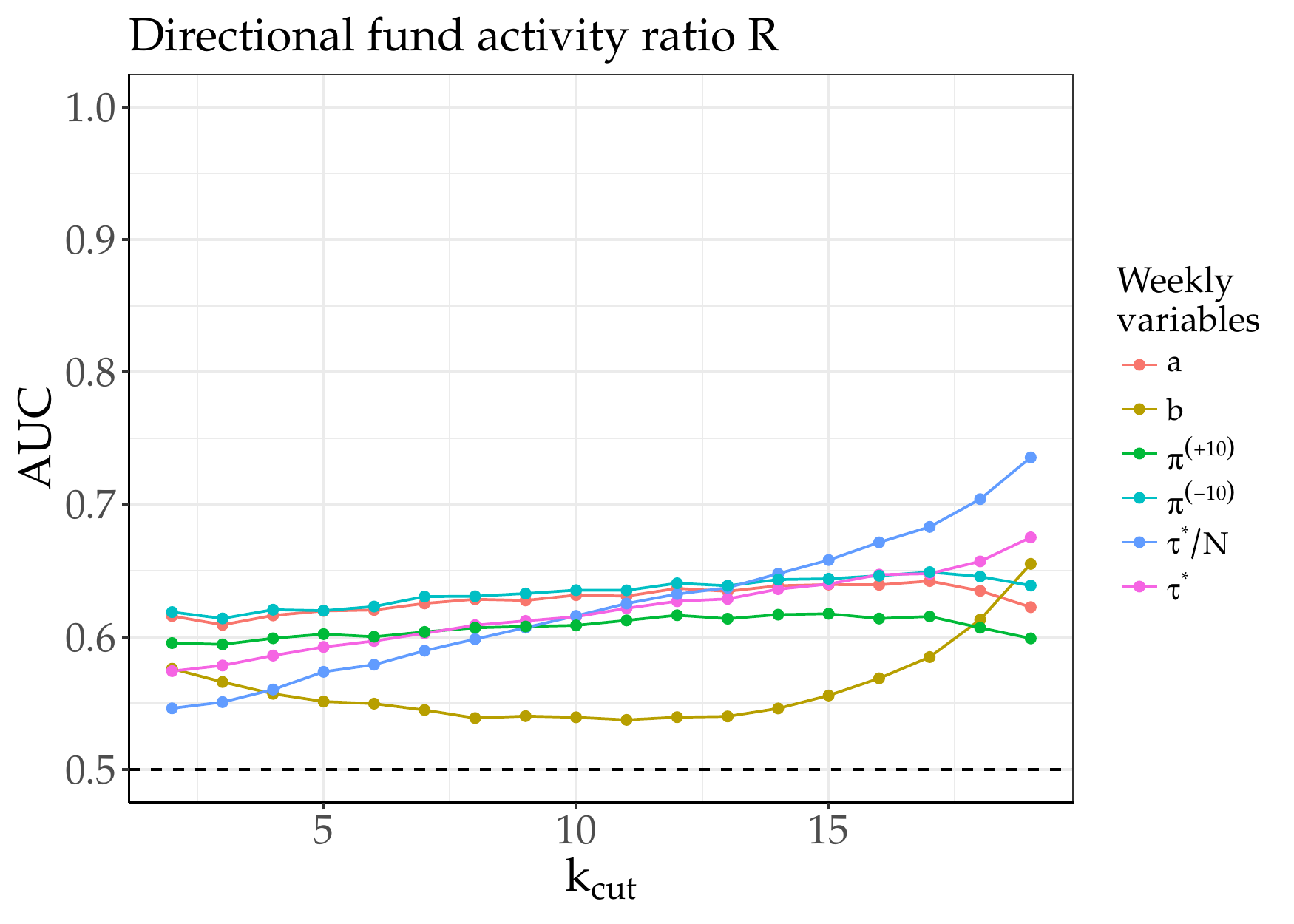}
\par\end{center}%
\end{minipage}
\par\end{centering}
\caption{Average quarterly Area Under Curve (AUC) corresponding to the classification
of funds belonging to quantile $k\ge k_{\text{cut}}$ for each market
order sign memory length measures. Left plot: averages of daily microstructural
quantities; right plot: averages of weekly microstructural quantities.
AUC averaged over the whole history.\label{fig:Area-under-curve}}

\end{figure}

Here, we focus on the following simple classification problem: for
each quarter, we split the 20 groups into two categories according
to quantile $k_{\text{cut}}$: assets belonging to the quantiles $k\le k_{\text{cut}}$
form the first category and the remaining ones the other one. Choosing
a memory length measure as the variable according to which one classifies
the assets during a given quarter, it is then straightforward to compute
the parametric Receiver Operating Characteristic (ROC) curve and its
associated area under curve (AUC) for a given quarter. Figure \ref{fig:Area-under-curve}
reports the AUC associated with $\pi^{(s10)}$, $a$, $b$, $\tau^{*}$
and $\tau^{*}/N$ for the quantiles of both $R$ and $S$, averaged
over the 32 quarters. One sees that $\pi^{(s10)}$ and $a$ are roughly
equivalent and are the best variables to discriminate the large values
of $R$ and $S$. Even more, their detection power with respect to
the quantiles of $S$ does not depend much on $k_{\text{cut}}$. 

\section{A theoretical insight}

\citet{lillo2005theory} establish the link between the size of meta-orders
and the order sign auto-correlation. In particular, they find that
if the distribution of the size $L$ of meta-orders, denoted by $P(L)$,
has a power-law tail $P(L)\propto L^{-(\beta+1)}$, then, assuming
that the sign order of each meta-order is equiprobably $-1$ or $+1$
and that there are exactly $M$ active meta-orders at any time for
a given asset, the auto-correlation function is given by $C(\tau)\simeq a\tau^{-b}=\frac{M^{\beta-2}}{\beta}\tau^{-(\beta-1)}$
for large $\tau$. This simple relationship gives two insights relevant
to our results. 

First, the pre-factor $a$ is a decreasing function of $M$ when $\beta<2$,
which is generally the case as $\beta\simeq1.5$ on average for example
in the London Stock Exchange \citep{lillo2005theory}. Identifying
$\beta-1$ with $b$ here shows that in our case $b<1$, hence that
$\beta<2$. This implies that the pre-factor $a$ is a proxy for the
number of meta-orders present in the market, \emph{ceteribus paribus},
which should then be strongly related to $S$. This is why the pre-factor
$a$ (and thus $\pi^{(10)}$) are among the best predictors of the
quantile range of $S$ (Fig.~ \ref{fig:auto_exps}).

Second, \textbf{$b$} being a proxy for an effective $\beta$, it
allows to gain some insight on the tail of $P(L)$. Since the measured
\textbf{$b$} decreases, the probability of very large meta-orders
increases, which is coherent with directional trading from large funds
after large price changes. Another explanation for the change of $b$
resides in the the way meta-orders are split: there must be feedback
loops between the exponent $\beta$ and the available liquidity which
in turns most probably depends on the current order flow, determined
in part by meta-orders.

\section{Concluding remarks}

This work has brought to light the fact that the influence of the
trading of large funds on the memory of market order signs is far
from negligible: large funds do not suppress long memory, but may
weaken it. When one knows \emph{ex postfacto} how large funds have
traded, there is a clear difference between limit order book dynamics
in which large fund took a substantial part and those barely touched
by them (relatively speaking). Reversely, even when averaging the
market order sign memory length measures over a quarter allows to
some extend to predict if large funds have been much involved in the
trading of a given asset.

The main limitation of the present work is the use of quarterly data
to characterise fund behaviour. Using labelled trades would open the
way to relate the properties of order book day by day and and to improve
our understanding of the link between the composition of meta-orders
and the memory length of market order signs.

\bibliographystyle{plainnat}
\bibliography{thesis}

\begin{thebibliography}{11}
\providecommand{\natexlab}[1]{#1}
\providecommand{\url}[1]{\texttt{#1}}
\expandafter\ifx\csname urlstyle\endcsname\relax
  \providecommand{\doi}[1]{doi: #1}\else
  \providecommand{\doi}{doi: \begingroup \urlstyle{rm}\Url}\fi

\bibitem[Blanc et~al.(2017)Blanc, Donier, and Bouchaud]{blanc2017quadratic}
Pierre Blanc, Jonathan Donier, and J-P Bouchaud.
\newblock Quadratic hawkes processes for financial prices.
\newblock \emph{Quantitative Finance}, 17\penalty0 (2):\penalty0 171--188,
  2017.

\bibitem[Bouchaud et~al.(2004)Bouchaud, Gefen, Potters, and
  Wyart]{bouchaud2004fluctuations}
Jean-Philippe Bouchaud, Yuval Gefen, Marc Potters, and Matthieu Wyart.
\newblock Fluctuations and response in financial markets: the subtle nature of
  'random' price changes.
\newblock \emph{Quantitative finance}, 4\penalty0 (2):\penalty0 176--190, 2004.

\bibitem[Challet et~al.(2016)Challet, Chicheportiche, Lallouache, and
  Kassibrakis]{challet2016trader}
Damien Challet, R{\'e}my Chicheportiche, Mehdi Lallouache, and Serge
  Kassibrakis.
\newblock Trader lead-lag networks and order flow prediction.
\newblock 2016.

\bibitem[Lillo(2007)]{lillo2007limit}
Fabrizio Lillo.
\newblock Limit order placement as an utility maximization problem and the
  origin of power law distribution of limit order prices.
\newblock \emph{The European Physical Journal B}, 55\penalty0 (4):\penalty0
  453--459, 2007.

\bibitem[Lillo and Farmer(2004)]{lillo2004long}
Fabrizio Lillo and J~Doyne Farmer.
\newblock The long memory of the efficient market.
\newblock \emph{Studies in nonlinear dynamics \& econometrics}, 8\penalty0 (3),
  2004.

\bibitem[Lillo et~al.(2005)Lillo, Mike, and Farmer]{lillo2005theory}
Fabrizio Lillo, Szabolcs Mike, and J~Doyne Farmer.
\newblock Theory for long memory in supply and demand.
\newblock \emph{Physical review e}, 71\penalty0 (6):\penalty0 066122, 2005.

\bibitem[Lynch and Zumbach(2003)]{zumbachlynch}
Paul Lynch and Gilles Zumbach.
\newblock Market heterogeneities and the causal structure of the volatility.
\newblock \emph{Quantitative Finance}, 3:\penalty0 320--331, 2003.

\bibitem[Toth et~al.(2015)Toth, Palit, Lillo, and Farmer]{toth2015equity}
Bence Toth, Imon Palit, Fabrizio Lillo, and J~Doyne Farmer.
\newblock Why is equity order flow so persistent?
\newblock \emph{Journal of Economic Dynamics and Control}, 51:\penalty0
  218--239, 2015.

\bibitem[Tumminello et~al.(2012)Tumminello, Lillo, Piilo, and
  Mantegna]{tumminello2011identification}
M.~Tumminello, F.~Lillo, J.~Piilo, and R.N. Mantegna.
\newblock Identification of clusters of investors from their real trading
  activity in a financial market.
\newblock \emph{New Journal of Physics}, 14:\penalty0 013041, 2012.

\bibitem[Zhou et~al.(2011)Zhou, Mu, Chen, and Sornette]{zhou2011investment}
Wei-Xing Zhou, Guo-Hua Mu, Wei Chen, and Didier Sornette.
\newblock Investment strategies used as spectroscopy of financial markets
  reveal new stylized facts.
\newblock \emph{PloS one}, 6\penalty0 (9):\penalty0 e24391, 2011.

\bibitem[Zumbach(2009)]{zumbach2009time}
Gilles Zumbach.
\newblock Time reversal invariance in finance.
\newblock \emph{Quantitative Finance}, 9\penalty0 (5):\penalty0 505--515, 2009.

\end{thebibliography}

\end{document}